\documentclass[preprintnumbers,amsmath,11pt,amssymb,floatfix,superscriptaddress,showpacs,nofootinbib]{revtex4}
\usepackage{graphicx}
\usepackage{epsfig}
\usepackage{bm}
\usepackage{amsfonts}

\begin{document}

\title{\Large Consequences of three modified forms of holographic dark energy models in bulk-brane interaction }
\medskip

\author{Antonio Pasqua}
\email{toto.pasqua@gmail.com} \affiliation{Department of Physics,University of Trieste, Trieste, Italy.}

\author{Surajit Chattopadhyay}
\email{surajit_2008@yahoo.co.in, surajcha@iucaa.ernet.in}
\affiliation{ Pailan College of Management and Technology, Bengal Pailan Park, Kolkata-700 104, India.}

\author{Ratbay Myrzakulov}
\email{rmyrzakulov@gmail.com}
\affiliation{Eurasian International Center for Theoretical Physics and Department of General and Theoretical Physics, Eurasian National University, Astana 010008, Kazakhstan}

\begin{abstract}
In this paper, we study the effects which are produced  by the interaction between a brane Universe and the bulk in which the Universe is embedded.
Taking into account the effects produced by the interaction between a brane Universe and the bulk, we  derived the Equation of State (EoS) parameter $\omega_D$ for
three different models of Dark Energy (DE), \emph{i.e.}  the Holographic DE (HDE) model with infrared (IR) cut-off given by the Granda-Oliveros cut-off, the Modified Holographic Ricci DE (MHRDE) model and a DE model which is function of the Hubble parameter $H$ squared and to higher derivatives of $H$. Moreover, we have considered two different cases of scale factor (namely, the power law and the emergent ones).  A nontrivial contribution of the DE is observed to be different from the standard matter fields confined to the brane. Such contribution has a monotonically decreasing behavior upon the evolution of the Universe for the emergent scenario of the scale factor, while monotonically  increasing for the power-law form of the scale factor $a(t)$.
\end{abstract}

\maketitle

\section{Introduction}
The evidence that our Universe is experiencing a phase of expansion with accelerated rate has been well demonstrated by cosmological data obtained from different independent observations of SNeIa, Cosmic Microwave Background Radiation (CMBR) anistropies, X-ray experiments and Large Scale Structures (LSS) \cite{1-1,1a,1b}.   Three main ideas have been suggested to give a reasonable explanation to the present day observed accelerated expansion of our Universe:
 the Cosmological Constant $\Lambda$ model, dark energy (DE) models and theories of modified gravity models. Thorough discussions on these three ideas are available in the reviews of \cite{Peebles,sahni1,copeland-2006,clift}. The Cosmological Constant $\Lambda$, which has EoS parameter $\omega = p/\rho = -1$, represents the earliest and the simplest theoretical candidate suggested in order to give a plausible explaination to the observational evidences of the Universe's present day accelerated expansion. It is well-known, anyway,  that there are two main problems associated with $\Lambda$: the fine-tuning and the cosmic coincidence problems. The former mainly asks why the vacuum energy density is so small (about an order of $10^{123}$ lower than what we can observe) while the latter asks why the  vacuum energy and DM give a nearly equal contribution at the present epoch even if they had an evolution which is independent and they had also evolved from  mass scales which are different (this fact represents a really strange coincidence if some internal connections between them are not taken into account). Till now, many attempts have been done in order to find a possible plausible explanation to the coincidence problem (see \cite{delcampo,delcampoa}).

The second idea suggested in order to possibly explain the observed accelerated expansion of the Universe involve DE models (reviewed in \cite{copeland-2006,bambaDE}). In relativistic cosmology, the cosmic acceleration we are able to observe can be described by the mean of a perfect fluid which pressure and energy density, indicated with $p$ and $\rho$, satisfy the relation given by $\rho + 3p < 0$. This kind of fluid is dubbed as Dark Energy (DE). The relation  $\rho + 3p < 0$ also tells us that the EoS parameter of the fluid $\omega$ must be in agreement with the condition $\omega  <-1/3$, while, from an observational point of view, it is a difficult work to constrain its exact value. The most direct evidence we have for the detection of DE is obtained from observations of supernovae of a type Ia (SNeIa) whose intrinsic luminosities can be safely considered practically uniform \cite{Peebles}.
If we assume that the DE idea is the right one in order to explain the present expansion of the Universe with accelerated rate, we must have that the largest amount of the total cosmic energy density $\rho_{tot}$ must be concentrated in the two Dark sectors, \emph{i.e.} Dark Energy (DE) and Dark Matter (DM) which represent, according to recent cosmological observations, about the 70$\%$ and about the 25$\%$ of the total energy density $\rho_{tot}$ of the present day Universe \cite{twothirds}. Moreover, the ordinary baryonic matter we are able to observe with our scientific  instruments contributes for only the 5$\%$ of $\rho_{tot}$. Furthermore, the radiation density gives a contribution to the total cosmic energy density which we can safely consider negligible. Many different models have been well studied in recent times to understand the exact nature of DE. Some of these models include tachyon, quintessence,  k-essence, quintom, Chaplygin gas, Agegraphic DE (ADE), NADE  and phantom. The various candidates of DE have been reviewed in \cite{copeland-2006,bambaDE}.

 A model of DE, motivated by the holographic principle, was proposed by Li \cite{3}  and it has been further studied in the references like \cite{3a,3b,4,5,5a,5b,6,rami1,rami2,rami3}. The energy density of HDE $\rho_D$ as follows:
\begin{eqnarray}
\rho_D = 3c^2 M_p^2 L^{-2}, \label{2}
\end{eqnarray}
with $c^2$ indicating a dimensionless constant parameter which which value $c$ is evinced by observational data:  for  a flat Universe (i.e. for $k=0$) it is obtained that $c=0.818_{-0.097}^{+0.113}$  and in the case of a
non-flat Universe (i.e. for $k=1$ or $k=-1$) it is obtained thar  $c=0.815_{-0.139}^{+0.179}$ \cite{n2primo,n2secondo}. Chen \emph{et al.} \cite{10} used the HDE model in order to drive inflation in the early evolutionary phases of the Universe. Jamil \emph{et al.} \cite{11} studied the EoS parameter $\omega_D$ of the HDE model considering not a constant but a time-dependent Newton's gravitational constant, \emph{i.e.} $G \equiv G\left( t \right)$; furthermore, they obtained that $\omega_D$ can be significantly modified in the low redshift limit.

Recently, the cosmic acceleration has been also well studied by imposing the concept of modification of gravity  \cite{nojo,nojo2}. This new model of gravity (predicted by string/M theory) gives a very natural gravitational alternative to the idea of the presence of exotic components. The explanation of the phantom, non-phantom and quintom phases of the Universe can be well described using modified gravity theories without the necessity of the introduction of a negative kinetic term in DE models. The relevance of modified gravity models for the late acceleration of the Universe has been recently studied by many Authors. Some of the most famous and known models of modified gravity are represented by braneworld models, $f\left(T\right)$ modified gravity (where $T$ indicates the torsion scalar), $f \left(R\right)$ modified gravity (where $R$ indicates the Ricci scalar curvature), $f \left(G\right)$ modified gravity (where $G$ indicates the Gauss-Bonnet invariant which is defined as $G=R^2-4R_{\mu \nu}R^{\mu \nu} + R_{\mu \nu \lambda \sigma}R^{\mu \nu \lambda \sigma}$, with $R$ representing the Ricci scalar curvature, $R_{\mu \nu}$ representing the Ricci curvature tensor and $R_{\mu \nu \lambda \sigma}$ representing the Riemann curvature tensor), $f \left(R,T\right)$  modified gravity,  $f \left(R,G\right)$  modified gravity, DGP  model, DBI  models, Horava-Lifshitz gravity and Brans-Dicke gravity. Modified theories of gravity have been reviewed in \cite{clift,mod1,mod2}.

Recently, the idea that our Universe is a brane which is embedded  in
a higher-dimensional space obtained a lot of attention from scientific community \cite{1,2ant,3bin,4-,5-,6-,7,sarida1}. The Friedmann equation on the brane
has some corrections with respect to the usual four-dimensional equation
\cite{4}. Binetruy \emph{et al.} \cite{3bin} found a term $H\propto \rho$,
which is problematic from an observational point of view. The model is consistent if  the tension on the brane and a
cosmological constant in the bulk are
considered. This leads to a cosmological version of the
Randall-Sundrum (RS) scenario of warped geometries \cite{4}. Bruck \emph{et al.} \cite{4} considered an interaction between the bulk and the
brane, which can be considered as another non-trivial aspect of braneworld
theories. The main aim of this paper is to disclose the
effects produced by the energy exchange between the brane and the bulk on
the evolutionary history of the Universe by taking into account the flow of energy
onto (or away) from the brane. In this paper, we will focus our attention to three particular DE models, i.e the HDE model with IR cut-off given by the recently proposed Granda-Oliveros cut-off, the Modified Holographic DE (MHRDE) model and a DE model which is proportional to the Hubble parameter $H$ squared and to higher time derivatives of $H$. Moreover, we will consider two different scale factors, \emph{i.e.} the power law and the emergent ones, in order to study the cosmological properties of the DE models in the Bulk-Brane interaction. Both the DE models and the scale factors considered will be described in details in the following Sections.
 This study is motivated by the works of \cite{setare1,sheykhi,Myung,kim}. In an interaction between the bulk and the brane,
Setare \cite{setare1} considered the holographic model of DE in
non-flat Universe under the assumption that the CDM energy density
on the brane is conserved while the HDE energy density
on the brane is not conserved because of to brane-bulk energy exchange.
Sheykhi \cite{sheykhi} considered the agegraphic models of DE in the framework of a braneworld scenario with brane-bulk energy exchange
under the assumption that the adiabatic equation for the DM is satisfied but it is violated for the Agegraphic DE (ADE) model because of the energy exchange between the brane and the bulk.
In the paper of Sheykhi \cite{sheykhi}, it was obtained that the
EoS parameter can evolve from the quintessence regime
to the phantom regime. Myung $\&$ Kim \cite{Myung} introduced the brane-bulk interaction in order to discuss a limitation of the cosmological Cardy-Verlinde formula which is useful for the holographic description of brane cosmology. They also showed that if there is presence of the brane-bulk interaction, it is not possible to derive the entropy representation of the
first Friedmann equation.

Saridakis \cite{sarida1} studied a generalized version of the HDE model arguing that it must be taked into account in the maximally subspace of a cosmological model; moreover he showed that, in the framework of brane cosmology, it leads to a bulk HDE which transfers its holographic nature to the effective $4D$ DE.
Furthermore, Saridakis \cite{sarida2} applied the bulk HDE in general $5D$ two-brane models and he also extracted the Friedmann equation on the physical brane, showing that in the general moving-brane case the effective $4D$ HDE has a quintom-like behavior for a large parameter-space area of a simple solution subclass.

In this paper, we consider an interaction between the bulk and the brane, which represents a non-trivial aspect of
the braneworld theories.  We also discuss the flow of energy onto or
away from the brane-Universe. We then apply this idea to a braneworld cosmology under the assumption that the DE energy density on
the brane is conserved, but the DE energy density on the brane is not conserved because of the brane-bulk energy exchange.

The plan of the paper is the following. In Section 2, we describe the main features of bulk-brane interaction. In Section 3, we describe the main features of the DE models considered in this paper; moreover, we derive the expression of the EoS parameter $\omega_D$ and the evolutionary form of the parameter $u$ (defined as $\frac{\chi}{\rho_m + \rho_D}$) for the DE models we are considering. In Section 4, we consider two different models of scale factors, (in particular, the power law and the emergent ones) in order to study the behavior of the expression of $\dot{u}$ derived in the previous Section. Finally, in Section 5, we write the Conclusion of this work.

\section{Bulk-brane energy exchange}
In this Section, we want to describe the main features of the bulk-brane interaction, introducing the main quantities useful for the following part of the work.\\
The bulk-brane action $S$ is given by the following expression \cite{setare1,cai}:
\begin{equation}\label{action}
S=\int d^5x\sqrt{-G}\left(\frac{R_5}{2\kappa_{5}^{2}}-\Lambda_5+L_{B}^{m}\right)+\int d^4x\sqrt{-g}(-\sigma+L_{b}^{m}),
\end{equation}
where $R_5$ represents the 5D curvature scalar, $\Lambda_5$ denotes the bulk cosmological constant, $\kappa_5$ stands for the 5D coupling constant, $\sigma$
indicates the brane tension, $G$ and $g$ denote  the determinant of the 5D and of the 4D metric tensors, respectively while $L_{B}^{m}$ and $L_{b}^{m}$  are the matter Lagrangian in the bulk and the matter Lagrangian in the brane.\\
We here consider the cosmological solution with a metric given by
\cite{setare1,cai}:
\begin{equation}\label{metric}
 ds^2=-n^2(t,y)dt^2+a^2(t,y)\gamma_{ij}dx^{i}dy^{j}+b^{2}(t,y)dy^{2},
\end{equation}
where $\gamma_{ij}$ represents the metric for the maximally symmetric three-dimensional space.
The non-zero components of Einstein tensor are given by \cite{setare1,cai}:
\begin{eqnarray}\label{g00}
  G_{00}&=&3\left\{\frac{\dot{a}}{a}\left(\dot{a}{a}+\dot{b}{b}\right)-\frac{n^2}{b^2}\left[\frac{a''}{a}\frac{a'}{a}
  \left(\frac{a'}{a}-\frac{b'}{b}\right)\right]+\frac{kn^2}{b^2}\right\}~,\\
  \label{gij}
  G_{ij}&=&\frac{a^{2}}{b^{2}}\gamma_{ij}\left[\frac{a'}{a}\left(\frac{a'}{a}+\frac{2n'}{n}\right)-
  \frac{b'}{b}\left(\frac{n'}{n}+\frac{2a'}{a}\right)+\frac{2a''}{a}+\frac{n''}{n}\right]+ \nonumber \\
  &&\frac{a^{2}}{n^{2}}\gamma_{ij}\left[\frac{\dot{a}}{a}\left(-\frac{\dot{a}}{a}+\frac{2\dot{n}}{n}\right)
  -\frac{2\ddot{a}}{a}+\frac{\dot{b}}{b}\left(-\frac{\dot{2a}}{a}+\frac{\dot{n}}{n}\right)
  -\frac{\ddot{b}}{b}\right]-k\gamma_{ij}~, \\
\label{g05}
 G_{05}&=&3\left(\frac{n'}{n}\frac{\dot{a}}{a}+\frac{a'}{a}\frac{\dot{b}}{b}-\frac{\dot{a}'}{a}\right)~,\\
\label{g55}
  G_{55}&=&3\left\{\frac{a'}{a}\left(\frac{a'}{a}+\frac{n'}{n}\right)-\frac{b^2}{n^2}
  \left[\frac{\dot{a}}{a}\left(\frac{\dot{a}}{a}-\frac{\dot{n}}{n}\right)+\frac{\ddot{a}}{a}\right]
  -\frac{kb^2}{a^2}\right\},
\end{eqnarray}
where $k$ denotes the curvature parameter of space which possible values are $k= 0, 1, -1$ which correspond, respectively, to a flat, a closed and an open Universe. Moreover, the primes and the dots indicate, respectively,  a derivative with respect to the variable $y$ and a derivative with respect to the variable $t$.
The 4D braneworld Universe is assumed to be at $y=0$. The
Einstein equations are given by:
\begin{eqnarray}
G_{\mu\nu}=\kappa_{5}^{2}T_{\mu\nu},
\end{eqnarray}
where we have that the stress-energy momentum tensor $T_{\mu\nu}$ has both bulk and brane components and it can be also  written as follows \cite{setare1,cai}:
\begin{equation}\label{stress}
  T^{\mu}_{\nu}=T^{\mu}_{\nu}|_{\sigma,b}+T^{\mu}_{\nu}|_{m,b}+T^{\mu}_{\nu}|_{\Lambda,B}+T^{\mu}_{\nu}|_{m,B},
\end{equation}
where:
\begin{eqnarray}\label{sigmab}
  T^{\mu}_{\nu}|_{\sigma,b}&=&\frac{\delta(y)}{b}\text{diag}(-\sigma,-\sigma,-\sigma,-\sigma,0),\\
\label{lambdab}
  T^{\mu}_{\nu}|_{\Lambda,B}&=&\text{diag}(-\Lambda_{5},-\Lambda_{5},-\Lambda_{5},-\Lambda_{5},-\Lambda_{5}),\\
\label{mb}
  T^{\mu}_{\nu}|_{m,b}&=&\frac{\delta(y)}{b}\text{diag}(-\rho,p,p,p,0),
\end{eqnarray}
where $p$ and $\rho$ represent, respectively, the total pressure and the total density on the brane.\\
By integrating Eqs. (\ref{g00}) and (\ref{gij}) with respect to the variable $y$ around the point $y = 0$ and assuming the $Z_2$ symmetry around the brane, we derive the following jump conditions:
\begin{eqnarray}\label{jump1}
 a'_{+}&=&-a'_{-}=-\frac{\kappa_{5}^{2}}{6}a_{0}b_{0}(\sigma+\rho),\\
\label{jump2}
 n'_{+}&=&-n'_{-}=\frac{\kappa_{5}^{2}}{6}b_{0}n_{0}(-\sigma+2\rho+3p).
\end{eqnarray}
The 2 subscripts + and - indicate, respectively, the sides corresponding to  $y > 0$ and $y < 0$, which represent the two sides of the brane embedded in the bulk. Moreover, the subscript 0 indicates quantities which are evaluated at $y = 0$.\\
Starting from the results of Eqs. (\ref{g05}) and (\ref{g55}), we can obtain the following expressions:
\begin{eqnarray}
\frac{n_0'\dot{a}_0}{n_0a_0}+ \frac{a_0'\dot{b}_0}{a_0b_0}-\frac{\dot{a}_0'}{a_0} &=& \frac{\kappa _5^2}{3}T_{05},\\
3\left\{\frac{a_0'}{a_0}\left(\frac{a_0'}{a_0}+\frac{n_0'}{n_0}\right)-\frac{b_0^2}{n_0^2}
  \left[\frac{\dot{a}_0}{a_0}\left(\frac{\dot{a}_0}{a_0}-\frac{\dot{n}_0}{n_0}\right)+\frac{\ddot{a}_0}{a_0}\right]
  -k\frac{b_0^2}{a_0^2}\right\} &=& -\kappa _5^2\Lambda_5 b_0^2 +\kappa_5^2T_{55},
\end{eqnarray}
where the terms $T_{05}$ and $T_{55}$ represent, respectively, the $05$ and $55$ components of $T_{\mu\nu}|_{m,b}$ when evaluated on the brane. \\
Moreover, using Eqs. (\ref{jump1}) and (\ref{jump2}), we obtain:
\begin{eqnarray}\label{field1}
  \dot{\rho}+3\frac{\dot{a}_{0}}{a_{0}}(\rho+p)&=&-\frac{2n_{0}^{2}}{b_{0}}T^{0}_{5},\\
\label{field2}
  \frac{1}{n_{0}^{2}}\left[\frac{\ddot{a}_{0}}{a_{0}}+\left(\frac{\dot{a}_{0}}{a_{0}}\right)^{2}
  -\frac{\dot{a}_{0}\dot{n}_{0}}{a_{0}n_{0}}\right]&+&\frac{k}{a_{0}^{2}}  \nonumber \\
  &=&\frac{\kappa_{5}^{2}}{3}\left(\Lambda_{5}+\frac{\kappa_{5}^{2}\sigma^{2}}{6}\right)  \nonumber \\
  &&-\frac{\kappa_{5}^{4}}{36}\left[\sigma(3p-\rho)+\rho(3p+\rho)\right]-\frac{\kappa_{5}^{2}}{3}T^{5}_{5}.
\end{eqnarray}
Considering an appropriate gauge with the coordinate frame $n_0 = b_0 = 1$, Eqs. (\ref{field1}) and (\ref{field2}) can be also expressed in the following equivalent forms:
\begin{eqnarray}
\dot{\rho} + 3H\left( 1+\omega  \right)\rho &=& -2 T_5^0, \\
\left(  \frac{\dot{a}}{a} \right)^2 &=& \Lambda -\frac{\kappa}{a^2} + \beta \rho^2 + 2\gamma \left(\rho + \chi  \right), \\
\dot{\chi}+4H\chi &=&
 2\left( \frac{\rho}{\sigma} +1 \right)T_5^0 - \frac{12}{\kappa_5^2}\frac{H}{\sigma}T_5^5,
\end{eqnarray}
where $\beta = \frac{\kappa_5^4}{36}$ and $\gamma = \frac{\sigma \kappa_5^4}{36}$. The effective 4D cosmological constant
$\Lambda$ on the brane, the bulk cosmological constant $\Lambda_5$, and the brane tension
$\sigma$ are well known to be constrained by the fine-tuning relation~\citep{maartens,bs1,Bazeia:2014tua,daRocha:2013ki}:
\begin{eqnarray}
\Lambda=\frac{\kappa_5^{2}}{2}\left(\frac{1}{6}\kappa_5^{2}\sigma^{2}+\Lambda_{5}\right).
\end{eqnarray}
 If we assume that the bulk matter (relative to bulk vacuum energy) is much less than the ratio of the brane matter to the brane vacuum energy, we can neglect the $T^{5}_{5}$ term: this can lead to the derivation of a solution that is largely independent of the bulk dynamics. If we take into account this approximation and we concentrate on the low-energy region with $\rho/\sigma\ll 1$, Eqs. (\ref{field1}) and (\ref{field2}) can be simplified, leading to the following system of equations:
\begin{eqnarray}
\label{field11} \dot{\rho}+3H(1+\omega)\rho &=&-2T^{0}_{5}=T, \\
\label{field22}  H^{2}&=&\frac{8 \pi G_{4}}{3}(\rho+\chi)-\frac{k}{a^{2}}+\Lambda, \\
\label{field33} \dot{\chi}&+&4H\chi\approx 2T^{0}_{5}=-T.
\end{eqnarray}
The auxiliary field $\chi$ (which appear in  Eqs. (\ref{field22}) and (\ref{field33})) incorporates non-trivial contributions of DE which differ from the standard matter fields confined to the brane. Hence, with the energy exchange $T$ between the bulk and brane, the usual energy conservation is violated.
We shall denote the energy density of DE by $\rho_D$. Since we will consider two dark components in the Universe, namely, DM and DE, we will have $\rho=\rho_D+\rho_{m}$.\\
In the following Section, three different DE models are concerned, namely, the HDE model with Granda-Oliveros cut-off, the MHRDE model and the DE model proportional to the Hubble parameter $H$ squared and to higher time derivatives of $H$ in the framework of bulk brane interaction. It is accomplished by using some of the concepts introduced in this Section and two choices of the scale factor, namely the power law and the emergent ones.

\section{MHRDE and GO DE MODEL in the bulk-brane interaction}
We now want to give a description of the DE models considered in this work and to find some relevant cosmological quantities. We will also introduce some relevant equations which will be useful for the understanding of the work.\\
The bulk-brane interaction has
been studied for various aspects, where in particular the effective DE of the braneworld Universe is dynamical, as a result of the non-minimal coupling, which gives a mechanism for bulk-brane interaction through gravity
\cite{setare1,setare2,cai}.  We assume here that the
adiabatic equation for the DM is satisfied, while it results to be
violated for DE due to the energy exchange between
the brane and the bulk \cite{setare1,cai}. Then, we obtain the following continuity equations:
\begin{eqnarray}\label{matter}
  \dot{\rho}_{m}&+&3H\rho_{m}=0,\\
  \label{energy}
  \dot{\rho}_D&+&3H(1+\omega_D)\rho_D=T.
\end{eqnarray}
We define the fractional energy densities for DM, DE and $\chi$, respectively, as follows:
\begin{eqnarray}
\Omega_m &=& \frac{\rho_m}{\rho_{cr}}, \label{fracden1}\\
\Omega_D &=& \frac{\rho_D}{\rho_{cr}},\label{fracden2}\\
\Omega_{\chi} &=& \frac{\chi}{\rho_{cr}},\label{fracden3} \\
\Omega_k &=& \frac{k}{a^2H^2}.\label{fracden4}
\end{eqnarray}
The Planck data provide the values $\Omega_m \approx 0.3089$ and $\Omega_D \approx 0.6911$ at $68\%$ CL \citep{adee}.
The critical energy density $\rho_{cr}$ (\emph{i.e.} the energy density required for flatness) is defined as follows:
\begin{eqnarray}
\rho_{cr} = \frac{3H^2}{8\pi G_4},
\end{eqnarray}
or, assuming units of $8\pi G_4=1$, as:
\begin{eqnarray}
\rho_{cr} = 3H^2. \label{rhocr}
\end{eqnarray}
Using the definition of $\rho_{cr}$ given in Eq. (\ref{rhocr}), we can write the fractional energy densities given in Eqs. (\ref{fracden1}),  (\ref{fracden2}) and   (\ref{fracden3}), respectively,      as follows:
\begin{eqnarray}\label{Fractional}
 \Omega_D&=&\frac{\rho_D}{3H^2}, \label{Fractional1}\\
 \Omega_{m}&=&\frac{\rho_{m}}{3H^2}, \label{Fractional2}\\
 \Omega_{\chi}&=&\frac{\chi}{3H^2}.\label{Fractional3}
\end{eqnarray}
The interaction between bulk and brane is given by the relation
$T=\Gamma \rho_D$, where the parameter $\Gamma$ represents the rate of
interaction.
 The Wilkinson Microwave
Anisotropy Probe (WMAP) satellite is well known to have measured the  curvature parameter $\Omega_k$ in Eq. (\ref{fracden4}), and, along with Baryon Acoustic Oscillation (BAO) and Hubble parameter measurement, it constrained the fractional energy density of the curvature parameter $k$ as $-0.0133 < \Omega_k < -0.0084$, in 95\% CL \citep{koma}. Eq. (\ref{fracden4}) for $\Omega_k$ is hence equal to  zero in this context.
Considering the parameter $u=\frac{\chi}{\rho_D+\rho_{m}}$,
the above equations lead to \cite{setare1}:
\begin{equation}\label{udot}
 \dot{u}=\left(\frac{3Hu\Omega_D}{\Omega_D+\Omega_{m}}\right)\left[\omega_D-
 \frac{1}{3}\left(\frac{\Omega_{m}}{\Omega_D}+1\right)-\frac{1+u}{u}\frac{\Gamma}{3H}\right].
\end{equation}
In this paper, we decided to consider the particular case corresponding to $\Lambda=0$. Furthermore,
following \cite{setare1}, we have chosen the following expression for $\Gamma$:
\begin{equation}\label{gamma}
 \Gamma=3b^2 (1+u)H,
\end{equation}
where $b^2$ represents a coupling parameter between DM and DE, also known as transfer strength \cite{q1,q1-4,q1-8,q1-9}.
From the observational data of the Gold SNeIa samples, CMBR data obtained from the WMAP and Planck satellites and the Baryonic Acoustic Oscillations (BAO) obtained thanks to the Sloan Digital Sky Survey (SDSS), the coupling parameter between DM and DE is estimated to assume a small positive value, satisfying the requirement for solving the cosmic coincidence problem and the constraints given by the second law of thermodynamics \cite{feng08}.
Cosmological observations of the CMBR anisotropies and of clusters of galaxies indicate that $b^2 < 0.025$  \cite{q4}. This evidence is in agreement with the fact that $b^2$ must be taken in the range of values [0,1] \cite{zhang-02-2006}, with $b^2 = 0$ representing the non-interacting FLRW model. \\
Using the definitions of the fractional energy densities given in Eqs. (\ref{Fractional1}), (\ref{Fractional2}) and (\ref{Fractional3}), we can rewrite the first Friedmann equation defined in Eq. (\ref{field22}) as follows:
\begin{eqnarray}
\Omega_m + \Omega_D + \Omega_{\chi}= 1.\label{allfrac}
\end{eqnarray}
Eq. (\ref{allfrac}) has the main property of relating all the fractional energy densities considered in this work.\\
Moreover, using Eqs. (\ref{Fractional1}), (\ref{Fractional2}) and (\ref{Fractional3}) along with the definition of $u$ and the relation $\rho_m + \rho_D = \Omega_m + \Omega_D$, we can easily obtain the following relation between the parameter $u$ and the fractional energy densities:
\begin{equation}
   u=\frac{1-\Omega_D-\Omega_{m}}{\Omega_D+\Omega_m}. \label{ufra}
\end{equation}
We now want to introduce three different energy density models for DE, \emph{i.e.} the HDE with Granda-Oliveros cut-off,  the Modified Holographic Ricci DE (MHRDE) model and the DE model proportional to $H^2$ and to higher  time derivatives of $H$. Before proceeding with calculations, we briefly describe these three models.\\
Recently,  Granda $\&$ Oliveros introduced a new IR cut-off based on purely dimensional ground which includes a term proportional to $\dot{H}$ and one term proportional to $H^2$. This new IR cut-off is known as Granda-Oliveros (GO) scale, indicated with the symbol $L_{GO}$ and it is given by \cite{grandaoliveros,grandaoliverosa}:
\begin{equation}
L_{GO}=\left( \alpha H^{2}+\beta \dot{H}\right) ^{-1/2},  \label{lgo5}
\end{equation}
where $\alpha $ and $\beta $ represent two constant parameters. In the limiting case corresponding to  $ \alpha = 2$ and $\beta = 1$, the GO scale $L_{GO}$ becomes proportional to the average radius of the Ricci scalar curvature (i.e., $L_{GO} \propto R^{-1/2}$) in the case the curvature parameter $k$ assume the value of zero (i.e. $k=0$), corresponding to a  flat Universe.
Recently, Wang $\&$ Xu \cite{wangalfa} have constrained the new HDE model in non-flat Universe using observational data. The best fit values of the two parameters $\left(\alpha, \beta   \right)$ with their confidence levels they found are given by $\alpha  = 0.8824^{+0.2180}_{-0.1163}(1\sigma)\,^{+0.2213}_{-0.1378}(2\sigma)$ and $\beta = 0.5016^{+0.0973}_{-0.0871}(1\sigma)\,^{+0.1247}_{-0.1102}(2\sigma)$ for non flat Universe, while for flat Universe they found that are $\alpha  = 0.8502^{+0.0984}_{-0.0875}(1\sigma)\,^{+0.1299}_{-0.1064}(2\sigma)$ and $\beta = 0.4817^{+0.0842}_{-0.0773}(1\sigma)\,^{+0.1176}_{-0.0955}(2\sigma)$.\\
We decided to consider the GO scale $L_{GO}$ as infrared cut-off for some specific reasons. If the IR cut-off chosen is given by the particle horizon, the HDE model cannot produce
an expansion of the Universe with accelerated rate. If we consider as cut-off of the system the future event horizon, the HDE model has a causality problem.
The DE models which consider the GO scale $L_{GO}$ depend only on local quantities, then it is possible to avoid the causality problem, moreover it is also possible to obtain the accelerated phase of the Universe.   \\
Granda $\&$ Oliveros considered that, since the origin of the HDE model is still not known exactly up to now, the consideration of the term with the time derivative of the Hubble parameter in the expression of the energy density of DE may be expected since this term appears in the curvature scalar and it has the right dimension. \\
The expression of the HDE energy density with $L_{GO}$ cut-off is given by:
\begin{equation}
\rho_{D_{GO}}= 3c^2\left( \alpha H^{2}+\beta \dot{H}\right).  \label{lgo5-1}
\end{equation}
We must underline here that we are considering the Planck mass $M_p$ equal to one. \\
Contrary to the HDE model based on the event horizon, the DE models which consider the GO scale depend only on local quantities, then it is possible to avoid in this way the causality problem.\\
The second model we consider in this paper is the Modified Holographic Ricci DE (MHRDE) model, which is given by the following expression:
\begin{equation}
\rho_{D_{MHRDE}}= \frac{2}{\alpha - \beta} \left(\dot{H} + \frac{3\alpha}{2}H^2    \right), \label{mhrde}
\end{equation}
where $\alpha$ and $\beta$ are the model parameters. Hereupon, we shall denote by $A_{\tiny\circ}$ any quantity $A_{MHRDE}$ related to the MHRDE model. This DE model was studied for the non-interacting case in reference \cite{chim1}, and Chimento \emph{et al.} have analyzed this this type of DE in interaction with DM as
Chaplygin gas \cite{mathew, chim2}.
In the limiting case corresponding to $\alpha = 4/3$ and $\beta =1$, the DE energy density model given in Eq. (\ref{mhrde}) leads to the DE energy density with Ricci scalar curvature for a spatially
at FLRW space-time as IR cut-off.\\
The use of the MHRDE is motivated by the holographic principle since we can relate the DE with an UV cut-off for the vacuum energy with an IR scale such as the one
given by the Ricci scalar curvature $R$. Moreover, it is possible to proceed in a different way taking into account that the Ricci scalar curvature $R$ is a new kind of DE, for example, a geometric DE instead of evoking the holographic principle. Irrespective of the origin
of the DE component, it modifies the Friedmann equation leading to a second order differential equation for the scale factor.\\
In this work, we decided to consider also a DE energy density model which was recently proposed by  Chen $\&$ Jing \cite{chens}. This new model is function of the Hubble parameter squared $H^2$ and of the first and second derivatives with respect to the cosmic time $t$ of the Hubble parameter $H$ and it is given by the following expression:
\begin{equation}
\rho_{D,higher}= 3c^2\left( \alpha H^{2}+\beta \dot{H} + \varrho \frac{\ddot{H}}{H}\right),  \label{lgo5-1higher}
\end{equation}
where $\alpha$, $\beta$ and $\gamma$ represent three arbitrary dimensionless parameters. The inverse of the Hubble parameter, \emph{i.e.} $H^{-1}$, is introduced in the first term of Eq. (\ref{lgo5-1higher}) so that the dimensions of each of the three terms are the same. \\
The behavior and the main cosmological features of the DE energy density model defined in Eq. (\ref{lgo5-1higher}) strongly depend on the three parameters of the model, \emph{i.e.} $\alpha$, $\beta$ and $\gamma$.
Eq. (\ref{lgo5-1higher}) can be considered as a generalization of two previously proposed energy density models of DE. In fact, in the limiting case corresponding to $\alpha = 0$, we recover the energy density of DE in the case the IR cut-off of the system is given by the Granda-Oliveros cut-off. Moreover, in the limiting case corresponding to $\alpha=0$, $\beta=1$ and $\gamma=2$, we obtain the expression of the energy density of DE with IR cut-off proportional to the average radius of the Ricci scalar  (i.e., $L \propto R^{-1/2}$) in the case of curvature parameter $k$ assumes the value of zero ($k=0$). \\
Using the expressions of the energy densities of DE given in Eqs. (\ref{lgo5-1}), (\ref{mhrde}) and  (\ref{lgo5-1higher}) in Eq. (\ref{Fractional1}), we obtain the following expressions for $\Omega_{D_{GO}}$, $\Omega_{D_{\tiny\circ}}$ and $\Omega_{D,higher}$:
\begin{eqnarray}
 \Omega_{D_{GO}}&=&\frac{\rho_{D_{GO}}}{3H^2}, \label{Fractional1GO}\\
  \Omega_{D_{\tiny\circ}}&=&\frac{\rho_{D_{\tiny\circ}}}{3H^2}, \label{Fractional1MH}\\
   \Omega_{D,higher}&=&\frac{\rho_{D,higher}}{3H^2}. \label{Fractional1higher}
\end{eqnarray}
The final expression of $\rho_m$ can be  derived by first solving the continuity equation for $\rho_m$ given in Eq. (\ref{matter}), yielding:
\begin{eqnarray}
\rho_m &=& \rho_{m0}a^{-3}, \label{rhoemme}
\end{eqnarray}
where  $\rho_{m0}$ indicates the present day of the energy density of DM.\\
Using the expression of $\rho_{m}$ given in Eq. (\ref{rhoemme}), we can write the fractional energy density of DM as follows:
\begin{eqnarray}
\Omega_m = \frac{\rho_{m0}a^{-3}}{3H^2}.
\end{eqnarray}
We now want to find the final expressions of the EoS parameter $\omega_D$ and of $\dot{u}$ for all the DE models considered in this work.\\
Differentiating Eq. (\ref{field22}) with respect to the cosmic time $t$ and using Eqs. (\ref{field11}) - (\ref{field33}), we obtain (considering units of $8\pi G_4=1$) the following expression of the time derivative of the Hubble parameter for flat Universe:
\begin{equation}\label{Hdot}
 \dot{H}=-\frac{1}{6}\left[3\rho_D(1+\omega_D)+3\rho_m+4\chi\right].
\end{equation}
Moreover, using Eqs. (\ref{field22}) and  (\ref{Hdot}) in Eqs. (\ref{lgo5-1}), (\ref{mhrde}) and (\ref{lgo5-1higher}), we obtain the following expressions for the EoS parameters of the DE models we are dealing with:
\begin{eqnarray}
\omega_{D_{GO}} &=&\frac{2}{3}\left( \frac{\alpha - 2\beta}{\beta}  \right)\frac{\chi}{\rho_{D_{GO}}} - \left(  1-\frac{2}{3}\frac{\alpha}{\beta} + \frac{2}{3c^2\beta}  \right) +\left( \frac{2\alpha - 3\beta}{3\beta}
\right)\frac{\rho_m}{\rho_{D_{GO}}}, \label{eosgo} \\
\omega_{D_{\tiny\circ}} &=& \left( \alpha -1  \right)\left(\frac{\rho_m}{\rho_{D_{\tiny\circ}}} \right)+\left(\alpha - \frac{4}{3} \right)\frac{\chi}{\rho_{D_{\tiny\circ}}} +\beta -1, \label{eosmh} \\
\omega_{D,higher} &=&\frac{2}{3}\left( \frac{\alpha - 2\beta}{\beta}  \right)\frac{\chi}{\rho_{D,higher}} - \left(  1-\frac{2}{3}\frac{\alpha}{\beta} + \frac{2}{3c^2\beta}  \right) +\left( \frac{2\alpha - 3\beta}{3\beta}
\right)\frac{\rho_m}{\rho_{D,higher}} \nonumber \\
 &&+ \frac{2\varrho}{\beta \rho_{D,higher}}\frac{\ddot{H}}{H} . \label{eoshigher}
\end{eqnarray}
Using the relation between $u$ and $\chi$ given by $u=\frac{\chi}{\rho_m + \rho_D}$, we can find the following expression for $\frac{\chi}{\rho_D}$:
\begin{eqnarray}
\frac{\chi}{\rho_D} = \frac{u\left( \rho_m + \rho_D  \right)}{\rho_D} = u\left( 1 + \frac{\rho_m}{\rho_D}   \right). \label{u}
\end{eqnarray}
Then, inserting Eq. (\ref{u}) in the expressions of the EoS parameters obtained in Eqs. (\ref{eosgo}), (\ref{eosmh}) and (\ref{eoshigher}), along with the relation $\frac{\rho_m}{\rho_D}=\frac{\Omega_m}{\Omega_D}$, we can rewrite Eqs. (\ref{eosgo}), (\ref{eosmh}) and (\ref{eoshigher}) as follows:
\begin{eqnarray}
\omega_{D_{GO}} &=&\frac{2u_{GO}}{3}\left( \frac{\alpha - 2\beta}{\beta}  \right)\left( 1 + \frac{\Omega_m}{\Omega_{D_{GO}}}   \right) - \left(  1-\frac{2}{3}\frac{\alpha}{\beta} + \frac{2}{3c^2\beta}  \right) +\left( \frac{2\alpha - 3\beta}{3\beta}  \right)\frac{\Omega_m}{\Omega_{D_{GO}}}, \label{eosgo1}\\
\omega_{D_{\tiny\circ}} &=& \left( \alpha -1  \right)\left(\frac{\Omega_m}{\Omega_{D_{\tiny\circ}}} \right)+\left(\alpha - \frac{4}{3} \right)u\left( 1 + \frac{\Omega_m}{\Omega_{D_{\tiny\circ}}}   \right) +\beta -1,\label{eosmh1}\\
\omega_{D,higher} &=&\frac{2u_{higher}}{3}\left( \frac{\alpha - 2\beta}{\beta}  \right)\left( 1 + \frac{\Omega_m}{\Omega_{D,higher}}   \right) - \left(  1-\frac{2}{3}\frac{\alpha}{\beta} + \frac{2}{3c^2\beta}  \right)   \nonumber\\
  &&+\left( \frac{2\alpha - 3\beta}{3\beta}  \right)\frac{\Omega_m}{\Omega_{D,higher}}+ \frac{2\varrho}{3\beta \Omega_{D,higher}}\frac{\ddot{H}}{H^3}  . \label{eoshigher1}
\end{eqnarray}
We must underline that in Eq. (\ref{eoshigher1}) we used the main definition of $\Omega_{D,higher}$ given in Eq. (\ref{Fractional1higher}).\\
Moreover, using the relation  $\Omega_D+\Omega_{m}=\left(1+u\right)^{-1}$ (which can be obtained from Eq. (\ref{ufra})) in Eqs. (\ref{eosgo1}), (\ref{eosmh1}) and (\ref{eoshigher1}), we can write:
\begin{eqnarray}
\omega_{D_{GO}} &=&\frac{2u_{GO}}{3}\left( \frac{\alpha - 2\beta}{\beta}  \right)\left[ \frac{1}{\left(1+u_{GO}\right)\Omega_{D_{GO}}}   \right] - \left(  1-\frac{2}{3}\frac{\alpha}{\beta} + \frac{2}{3c^2\beta}  \right) \nonumber\\
 && + \left( \frac{2\alpha -3\beta}{3\beta}  \right)\left[ \frac{1}{\left(1+u_{GO}\right)\Omega_{D_{GO}}}  -1 \right]\nonumber \\
&=& \left[\frac{2u_{GO}}{3}\left( \frac{\alpha - 2\beta}{\beta} \right) + \frac{2}{3}\frac{\alpha}{\beta}-1 \right]\left[ \frac{1}{\left(1+u_{GO}\right)\Omega_{D_{GO}}}   \right] - \frac{2}{3c^2\beta}, \label{eosgo2}\\
\omega_{D_{\tiny\circ}} &=& \left( \alpha -1  \right)\left[\frac{1}{\left(1+u_{\tiny\circ}\right)\Omega_{D_{\tiny\circ}}-1} \right]+\left(\alpha - \frac{4}{3} \right)u\left( 1 + \frac{\Omega_m}{\Omega_{D_{\tiny\circ}}}   \right) +\beta -1,\label{eosmh2}\\
\omega_{D,higher} &=&\frac{2u_{higher}}{3}\left( \frac{\alpha - 2\beta}{\beta}  \right)\left[ \frac{1}{\left(1+u_{higher}\right)\Omega_{D,higher}}   \right] - \left(  1-\frac{2}{3}\frac{\alpha}{\beta} + \frac{2}{3c^2\beta}  \right) \nonumber\\
 && + \left( \frac{2\alpha -3\beta}{3\beta}  \right)\left[ \frac{1}{\left(1+u_{higher}\right)\Omega_{D,higher}}  -1 \right]  + \frac{2\varrho}{3\beta \Omega_{D,higher}}\frac{\ddot{H}}{H^3}\nonumber \\
&=& \left[\frac{2u_{higher}}{3}\left( \frac{\alpha - 2\beta}{\beta} \right) + \frac{2}{3}\frac{\alpha}{\beta}-1 \right]\left[ \frac{1}{\left(1+u_{higher}\right)\Omega_{D,higher}}   \right] - \frac{2}{3c^2\beta} \nonumber \\
&&+ \frac{2\varrho}{3\beta \Omega_{D,higher}}\frac{\ddot{H}}{H^3}. \label{eoshigher2}
\end{eqnarray}
Using Eqs. (\ref{matter}) and (\ref{energy}) along with the expression of $\Gamma$ we have chosen, we obtain the following expression for the time evolution of $u$ for the three different DE models we are dealing with:
\begin{eqnarray}
 \dot{u}_{GO}&=&\left(\frac{3Hu_{GO}\Omega_{D_{GO}}}{\Omega_{D_{GO}}+\Omega_{m}}\right)\left[\omega_{D_{GO}}-
 \frac{1}{3}\left(\frac{\Omega_{m} + \Omega_{D_{GO}}}{\Omega_{D_{GO}}}\right)-\frac{b^2\left(1+u_{GO}\right)^2}{u_{GO}}\right],    \label{uasa1}\\
 \dot{u}_{\tiny\circ}&=&\left(\frac{3Hu_{\tiny\circ}\Omega_{D_{\tiny\circ}}}{\Omega_{D_{\tiny\circ}}+\Omega_{m}}\right)\left[\omega_{D_{\tiny\circ}}-
 \frac{1}{3}\left(\frac{\Omega_{m} + \Omega_{D_{\tiny\circ}}}{\Omega_{D_{\tiny\circ}}}\right) -\frac{b^2\left(1+u_{\tiny\circ}\right)^2}{u_{\tiny\circ}}\right],    \label{uasa1-1}\\
  \dot{u}_{higher}&=&\left(\frac{3Hu_{higher}\Omega_{D,higher}}{\Omega_{D,higher}+\Omega_{m}}\right)\left[\omega_{D,higher}-
 \frac{1}{3}\left(\frac{\Omega_{m} + \Omega_{D,higher}}{\Omega_{D,higher}}\right)-\frac{b^2\left(1+u_{higher}\right)^2}{u_{higher}} \right].    \label{uasahigher}
\end{eqnarray}
Inserting the expressions of the EoS parameters obtained in Eqs. (\ref{eosgo2}), (\ref{eosmh2}) and (\ref{eoshigher2}) into Eqs. (\ref{uasa1}), (\ref{uasa1-1}) and (\ref{uasahigher}) and using the relation $\Omega_D+\Omega_{m}=(1+u)^{-1}$, we obtain the following expressions for the three different DE models considered:
\begin{eqnarray}
 \dot{u}_{GO} &=& \frac{3Hu_{GO}\left(  1+u_{GO} \right)}{\Omega_{D_{GO}}}  \left\{  \left[ \frac{2}{3}u_{GO}\left( \frac{\alpha - 2\beta}{\beta}  \right)  + \frac{2\alpha}{3\beta}-\frac{4}{3}   \right]\left[  \frac{1}{\left(1+u_{GO}\right)\Omega_{D_{GO}}}\right]\right. \nonumber \\
&&\left.  -\frac{2}{3c^2\beta}  -\frac{b^2\left(1+u_{GO}\right)^2}{u_{GO}}\right\} ,  \label{uasanew1}\\
 \dot{u} _{\tiny\circ}&=&  \frac{3Hu_{\tiny\circ}\left(  1+u_{\tiny\circ} \right)}{\Omega_{D_{\tiny\circ}}}  \left\{ \left(\alpha - \frac{4}{3}\right) \left(  \frac{1}{\Omega_{D_{\tiny\circ}}}\right) +\beta - \alpha  -\frac{b^2\left(1+u_{\tiny\circ}\right)^2}{u_{\tiny\circ}} \right\}, \label{uasanew2}\\
  \dot{u}_{higher} &=& \frac{3Hu_{higher}\left(  1+u_{higher} \right)}{\Omega_{D,higher}}  \left\{  \left[ \frac{2}{3}u_{higher}\left( \frac{\alpha - 2\beta}{\beta}  \right)  + \frac{2\alpha}{3\beta}-\frac{4}{3}   \right]\times \right.\nonumber \\
 &&\left. \left[  \frac{1}{\left(1+u_{higher}\right)\Omega_{D,higher}}\right] + \frac{2\varrho}{3\beta \Omega_{D,higher}}\frac{\ddot{H}}{H^3}\right. \nonumber \\
&&\left.  -\frac{2}{3c^2\beta}  -\frac{b^2\left(1+u_{higher}\right)^2}{u_{higher}}\right\} .  \label{uasanew3}
\end{eqnarray}

In the following Section, we will study the behavior of the evolutionary forms of  $\dot{u}_{GO}$, $\dot{u}_{\tiny\circ}$ and $\dot{u}_{higher}$ obtained, respectively, in Eqs. (\ref{uasanew1}), (\ref{uasanew2}) and (\ref{uasanew3}) for two different choices of the scale factor, \emph{i.e.} the power law and the emergent ones. Using the reconstructed expressions of $u$, we will use them in order to study the behavior of the EoS parameters for the three DE models we are considering and obtained, respectively, in Eqs. (\ref{eosgo1}), (\ref{eosmh1}) and (\ref{eoshigher1}).
We must also emphasize that will find the final expression of the term $\frac{\ddot{H}}{\Omega_D H^3}$ according to the choice of the scale factor we will make.\\

\section{Scale factors}
In this Section,  we want to study the behavior of the reconstructed expressions of $u$, determined from $\dot{u}_{GO}$, $\dot{u}_{\tiny\circ}$ and $\dot{u}_{higher}$ obtained, respectively, in Eqs. (\ref{uasanew1}),  (\ref{uasanew2}) and  (\ref{uasanew3}), for two different choices of the scale factor, \emph{i.e.} the power law and the emergent ones.\\
In order to find the final expressions of $\dot{u}$ for the different choices of scale factor, we need to calculate the expressions of $\Omega_{D_{GO}}$, $\Omega_{D_{\tiny\circ}}$ and $\Omega_{D,higher}$ (defined, respectively, in Eqs. (\ref{Fractional1GO}), (\ref{Fractional1MH}) and (\ref{Fractional1higher})) and $H$ for the relevant case of the scale factor (remembering that $H=\frac{\dot{a}}{a}$).  We will then plot the reconstructed expressions of $u$ derived from $\dot{u}$ we obtained for some range of values of the parameters involved. Thanks to the reconstructed expression of $u$, we can plot the behavior of the EoS parameter $\omega_D$ for the relevant model and the specific scale factor.

\subsection{Power Law form of the scale factor}
We start the study of the different scale factors taking into account the power law scenario. \\
Following Setare \cite{set28}, we consider the power law case of the scale factor in the following form:
\begin{eqnarray}
a(t) = a_0\left(t_s -t\right)^n, \label{plscale}
\end{eqnarray}
where $a_0$, $t_s$ and $n$ are three constants. The term $t_s$ indicates the finite future singularity
time and the scale factor defined in Eq. (\ref{plscale}) is often used in scientific literature in order to check the type II (sudden singularity) or type IV (which corresponds to $\dot{H}$) for positive values of the power law index $n$. \\
We have that the derivative of the scale factor given in Eq. (\ref{plscale}) with respect to the cosmic time $t$ is given by:
\begin{eqnarray}
\dot{a}\left(t \right) = -na_0\left(t_s -t\right)^{n-1}. \label{plscale2}
\end{eqnarray}
Using the results of Eqs. (\ref{plscale}) and (\ref{plscale2}), we obtain that the expression of the Hubble parameter and its first and second time derivatives  are given, respectively, by:
\begin{eqnarray}
H &=&  \frac{\dot{a}}{a}=-\frac{n}{t_s -t}, \label{Hpl} \\
\dot{H}  &=& \frac{\dot{H}}{dt}=-\frac{n}{\left(t_s -t\right)^2}, \label{Hdotpl}\\
\ddot{H} &=&\frac{\ddot{H}}{dt^2}=-\frac{2n}{\left(t_s -t\right)^3} \label{Hddotpl}.
\end{eqnarray}

\begin{figure}[ht]
\centering\includegraphics[width=8cm]{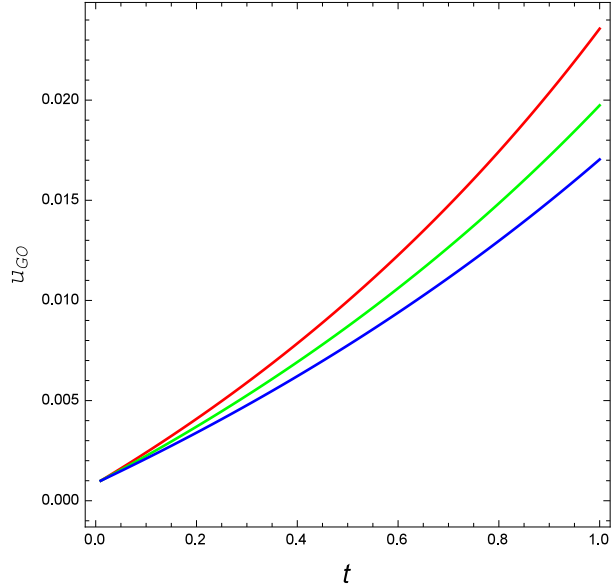}
\caption{Plot of $u_{GO}$ for power-law scale factor against the time $t$. The increasing pattern indicates that the non-trivial contribution of DE increases with the evolution of the Universe.}
\label{ugopowerlaw}
\end{figure}

\begin{figure}[ht]
\centering\includegraphics[width=8cm]{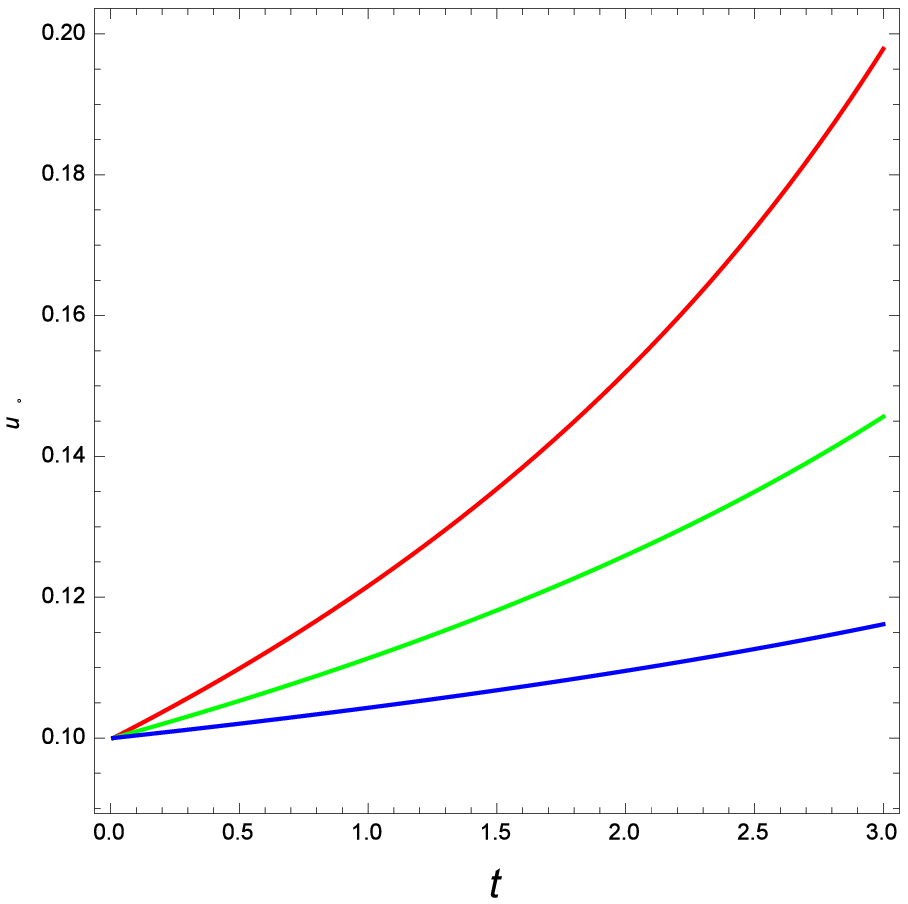}
\caption{Plot of $u_{\tiny\circ}$ for power-law scale factor against the time $t$. The increasing pattern indicates the non-trivial contribution of DE increases with the evolution of the Universe.} \label{umh}
\end{figure}

\begin{figure}[ht]
\centering\includegraphics[width=8cm]{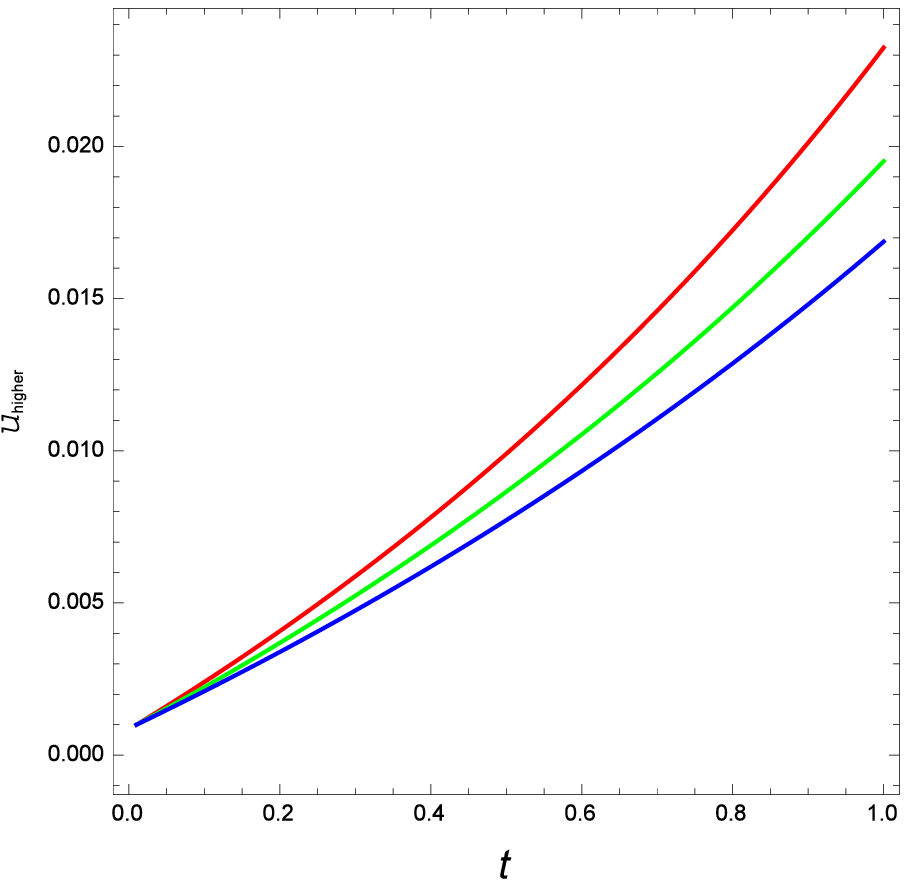}
\caption{Plot of $u_{higher}$ for power-law scale factor against the time $t$. The increasing pattern indicates that the non-trivial contribution of DE increases with the evolution of the Universe.}
\label{uhigherpowerlaw}
\end{figure}

Using the expression of $H$ obtained in Eq. (\ref{Hpl}) and the expressions of $\Omega_{D_{GO}}$,  $\Omega_{D_{\tiny\circ}}$ and $\Omega_{higher}$, obtained inserting in Eqs.  (\ref{Fractional1GO}), (\ref{Fractional1MH}) and (\ref{Fractional1higher}) the expressions of $\rho_{D_{GO}}$, $\rho_{D_{\tiny\circ}}$ and $\rho_{higher}$ defined in Eqs. (\ref{lgo5-1}), (\ref{mhrde}) and (\ref{lgo5-1higher}) calculated for  $H$, $\dot{H}$ and $\ddot{H}$ given in Eqs.  (\ref{Hpl}), (\ref{Hdotpl}) and (\ref{Hddotpl}), we derive the following expressions for $\dot{u}_{GO}$, $\dot{u}_{\tiny\circ}$ and $u_{higher}$:
\begin{eqnarray}
 \dot{u}_{GO} &=& 3u_{GO}\left(  1+u_{GO} \right) \left[  \frac{n^2}{c^2 (t_s-t) (\alpha -n \beta )}  \right]\times \nonumber \\
  &&\left\{  \left[ \frac{2}{3}u\left( \frac{\alpha - 2\beta}{\beta}  \right)  + \frac{2\alpha}{3\beta}-\frac{4}{3}   \right] \frac{n}{c^2 (-\alpha +n \beta )\left(1+u_{GO}\right)} \right. \nonumber \\
&&\left.   -\frac{2}{3c^2\beta}  -\frac{b^2\left(1+u_{GO}\right)^2}{u_{GO}}\right\} ,  \label{ugopl}\\
 \dot{u}_{\tiny\circ} &=&  3u_{\tiny\circ}\left(  1+u_{\tiny\circ} \right) \left \{  \frac{3n^2\left( -\alpha + \beta  \right)}{\left( t_s -t  \right)\left( 3n\alpha -2  \right)}  \right\} \times \nonumber \\
 &&\left\{ \left[\left(\alpha - \frac{4}{3}\right) \frac{3n (\alpha -\beta )}{-2+3 n \alpha } \right]+\beta - \alpha  -\frac{b^2\left(1+u_{\tiny\circ}\right)^2}{u_{\tiny\circ}} \right\}, \label{umhpl}\\
  \dot{u}_{higher} &=& 3u_{higher}\left(  1+u_{higher} \right) \left[  \frac{n^2}{c^2 (t_s-t) (\alpha -n \beta )}  \right]\times \nonumber \\
  &&\left\{  \left[ \frac{2}{3}u\left( \frac{\alpha - 2\beta}{\beta}  \right)  + \frac{2\alpha}{3\beta}-\frac{4}{3}   \right] \frac{n}{c^2 (-\alpha +n \beta )\left(1+u_{higher}\right)} \right. \nonumber \\
&&\left. + \frac{4\varrho}{3c^2\beta \left[ n\left(n\alpha - \beta\right) + 2\varrho  \right]}  -\frac{2}{3c^2\beta}  -\frac{b^2\left(1+u_{higher}\right)^2}{u_{higher}}\right\} .  \label{uhigherpl}
\end{eqnarray}
By using numerical integration, the evolution for  $u_{GO}$, $u_{\tiny\circ}$ and $u_{higher}$ are depicted in Figures  \ref{ugopowerlaw},  \ref{umh} and \ref{uhigherpowerlaw}. For the case pertaining to the HDE model with GO cut-off, we  considered three different cases, \emph{i.e.} for $\beta = 4.4$ (plotted in red), $\beta= 4.6$    (plotted in green) and $\beta= 4.8$    (plotted in blue), while the other parameters involved have been chosen as    $\alpha =4$, $n=1.4$, $c^2=0.818$, $b^2 = 0.025$ and $t_s=7$. It is worthwhile to emphasize that $u_{GO}$ has a monotonically increasing behavior for all the three cases considered. \\
For the case corresponding to the MHRDE model, three different cases are regarded, namely $\beta= 2.5$ (plotted in red),  $\beta=3$   (plotted in green) and $\beta= 3.5$  (plotted in blue), while the other parameters involved have been chosen as $\alpha = 4$, $n= 1.4$, $c^2=0.818 $, $b^2 =0.025 $ and $t_s=7$. As for the previous case, an increasing profile of $u_{\tiny\circ}$ can be observed,  for all the three cases considered. \\
For the model proportional to higher time derivatives of the Hubble parameter $H$, we have considered three different cases, corresponding to $\beta = 4.4$ (plotted in red), $\beta= 4.6$    (plotted in green) and $\beta= 4.8$    (plotted in blue), while the other parameters involved have been chosen as    $\alpha =4$, $\varrho = 5$, $n=1.4$, $c^2=0.818$, $b^2 = 0.025$ and $t_s=7$. We can observe in Figure \ref{uhigherpowerlaw} that $u_{higher}$ monotonically increases for all the cases considered, as also found for the other two DE model considered.\\
These increasing behaviors of $u_{GO}$, $u_{\tiny\circ}$ and $u_{higher}$ shown in Figures \ref{ugopowerlaw}, \ref{umh} and \ref{uhigherpowerlaw} clearly  indicate a non-trivial contribution of DE, contribution which increases with the temporal evolution of the Universe.

\begin{figure}[ht]
\centering\includegraphics[width=8cm]{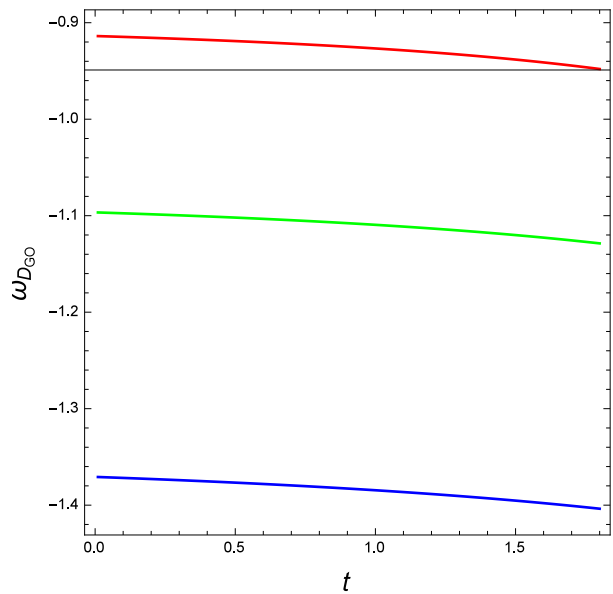}
\caption{Plot of $\omega_{D_{GO}}$ against the time $t$ for power-law scale factor. We observe a decreasing behavior for all cases considered.For $\beta = 4.4$ (plotted in red), $\omega_{D_{GO}}$ starts being $>-1$, then it decreases and it can eventually cross $\omega_D =-1$. For the other two cases, we obtain that $\omega_{D_{GO}}$ is always lower that $-1$. } \label{omegagopowerlaw}
\end{figure}

\begin{figure}[ht]
\centering\includegraphics[width=8cm]{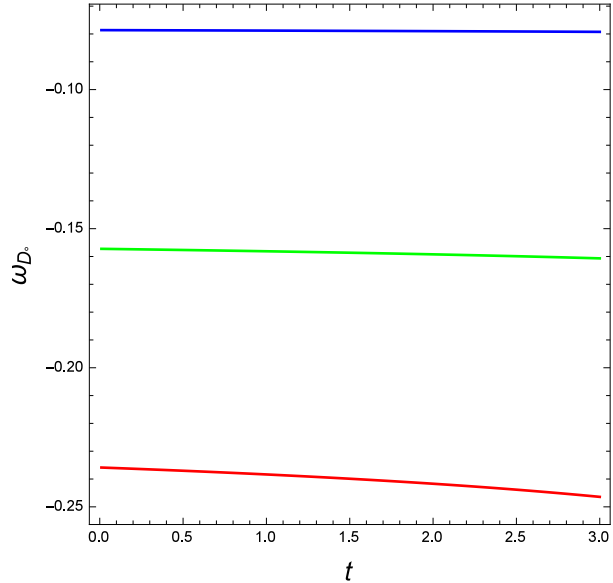}
\caption{Plot of $\omega_{D_{\tiny\circ}}$ against the time $t$ for the emergent scale factor. For all the cases considered, we have that $\omega_{D_{\tiny\circ}}$ has a slowly decreasing pattern and it is always greater than $-1$.} \label{omegamhrdepowerlaw}
\end{figure}

\begin{figure}[ht]
\centering\includegraphics[width=8cm]{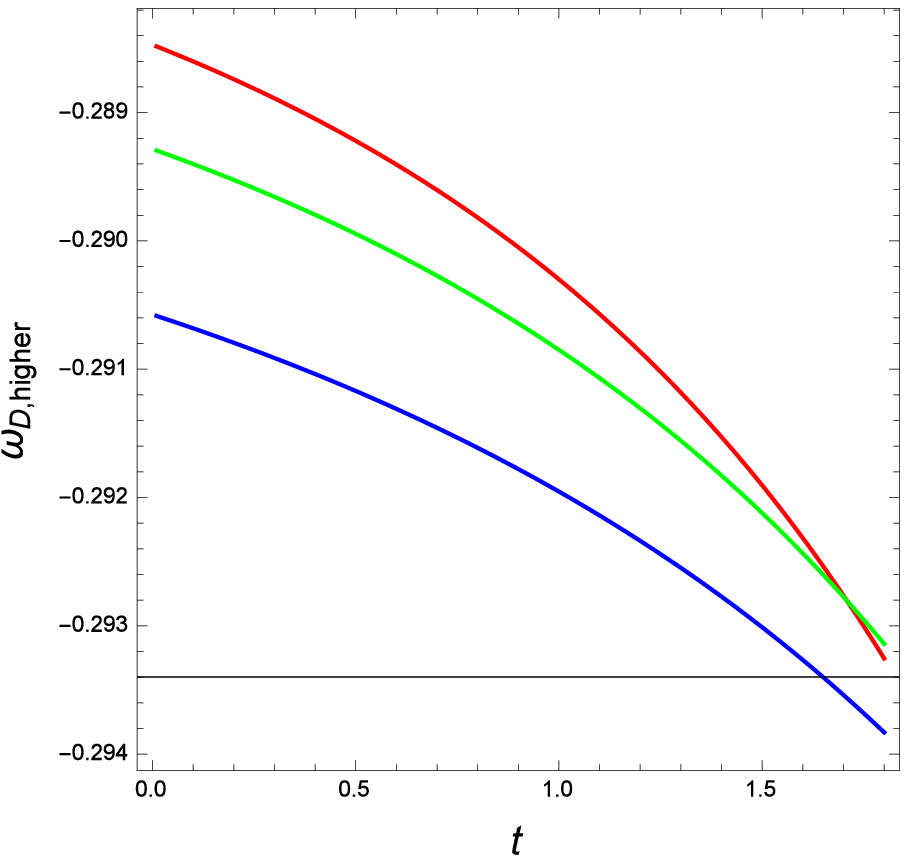}
\caption{Plot of $\omega_{D,higher}$ against the time $t$ for power-law scale factor. For all the cases considered, we have that $\omega_{D,higher}$ has a decreasing pattern and it is always greater than $-1$. } \label{omegahigherpowerlaw}
\end{figure}

Using the reconstructed expressions of $u_{GO}$, $u_{\tiny\circ}$ and $u_{higher}$  obtained, respectively, from Eqs. (\ref{ugopl}), (\ref{umhpl}) and (\ref{uhigherpl}) and plotted in Figures \ref{ugopowerlaw}, \ref{umh} and \ref{uhigherpowerlaw}, we can also derive and plot the profile of the EoS parameters obtained in Eqs. (\ref{eosgo1}), (\ref{eosmh1}) and (\ref{eoshigher1}) for the three DE models concerned.\\
For the DE model with GO cut-off, we obtain that, for $\beta = 4.4$, the EoS parameter $\omega_{D_{GO}}$ starts being $>-1$, while with the passing of the time, it decreases and it asymptotically reaches the value $-1$ and can eventually cross it. For the other two cases, we obtain that $\omega_{D_{GO}}$ has a decreasing behavior, being always lower that $-1$.\\
Instead, for the MHRDE model, we obtain that the EoS parameter $\omega_{D_{\tiny\circ}}$ has a slowly decreasing behavior for all the three cases considered, staying always greater than $-1$.\\
For the model proportional to higher time derivatives of the Hubble parameter $H$, we observe a slowly decreasing behavior of the EoS parameter $\omega_{D,higher}$, with $\omega_{D,higher}>-1$ for the range of time considered. It is possible that, for sufficiently high time, the three models can cross the value $\omega_D=-1$.\\

\begin{figure}[ht]
\centering\includegraphics[width=8cm]{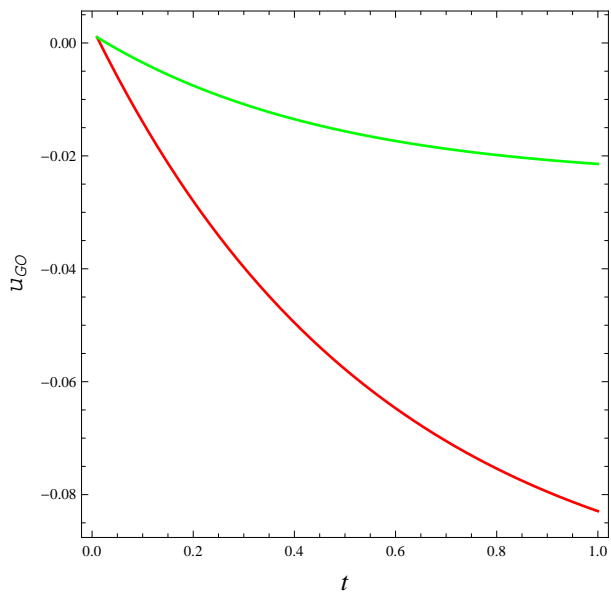}
\caption{Plot of $u_{GO}$ for power-law scale factor against the time $t$ for the limiting cases of $\alpha = 2$ and $\beta =1$ (plotted in   green) and $\alpha = 0.8502$ and $\beta = 0.4817$ (plotted in   red). }
\label{ugopowerlawlim}
\end{figure}

We now consider some particular values of the parameters involved.\\
For the DE model with GO cut-off, we study the case corresponding to the Ricci scale, which is recovered for $\alpha = 2$ and $\beta =1$ (plotted in  green) and we will also consider the case corresponding to $\alpha = 0.8502$ and $\beta = 0.4817$ (plotted in red). Instead, for the MHRDE model, we will consider the case corresponding to the Ricci scale, which is recovered for $\alpha = 4/3$ and $\beta =1$. The values of the other parameters have been taken as the previous cases considered. \\
We can clearly see in Figure \ref{ugopowerlawlim} that, for both limiting cases, $u_{GO}$ has a decreasing behavior while $\omega_{D_{GO}}$ has a slowly increasing behavior. Moreover, we have that for the case pertaining to the Ricci scale,   $\omega_{D_{GO}}$ is always greater than -1 while for the case with $\alpha = 0.8502$ and $\beta = 0.4817$ it is always lower than -1.\\
For the limiting case of the MHRDE, we observe that $u_{\tiny\circ}$  has an increasing behavior  while $\omega_{D_{\tiny\circ}}$ slowly decreases, being always greater than -1.

\begin{figure}[ht]
\centering\includegraphics[width=8cm]{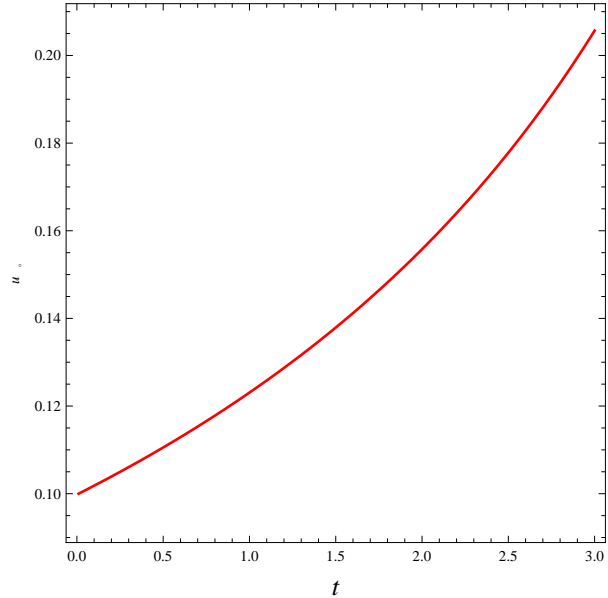}
\caption{Plot of $u_{\tiny\circ}$ for power-law scale factor against the time $t$ for the limiting case of $\alpha = 4/3$ and $\beta =1$. } \label{umhlim}
\end{figure}

\begin{figure}[ht]
\centering\includegraphics[width=8cm]{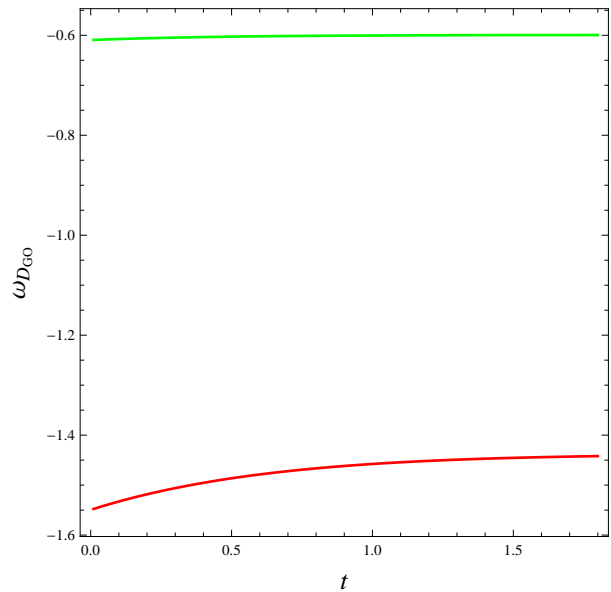}
\caption{Plot of $\omega_{D_{GO}}$ against the time $t$ for power-law scale factor for the limiting cases of $\alpha = 2$ and $\beta =1$ (plotted in   green) and $\alpha = 0.8502$ and $\beta = 0.4817$ (plotted in   red).} \label{omegagopowerlawlim}
\end{figure}

\begin{figure}[ht]
\centering\includegraphics[width=8cm]{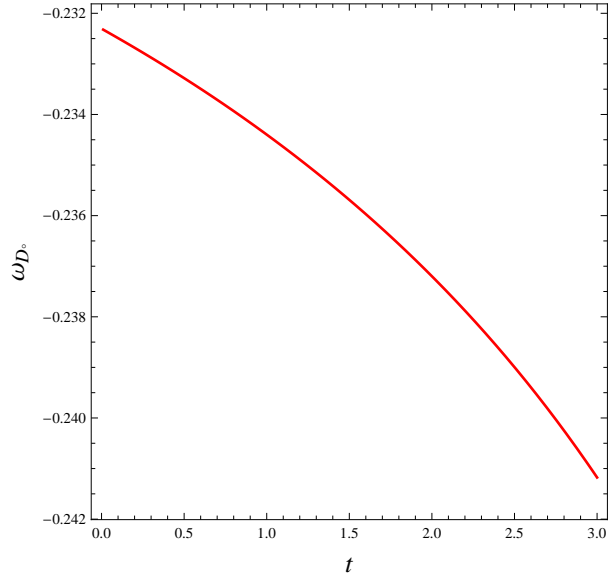}
\caption{Plot of $\omega_{D_{\tiny\circ}}$ against the time $t$ for the emergent scale factor for the limiting case of $\alpha = 4/3$ and $\beta =1$. } \label{omegamhrdepowerlawlim}
\end{figure}

\subsection{Scale factor pertaining to emergent scenario}
We now consider the second scale factor chosen in this work, \emph{i.e.} the emergent one.\\
This form of scale factor $a\left( t\right)$, as stated in \cite{debnachatto,ghoshchatto,debna} is given by:
\begin{eqnarray}
a\left( t \right)= a_0 \left(e^{\mu t}+\lambda \right)^m, \label{aeme}
\end{eqnarray}
where $a_0$, $\lambda$ , $\mu$ and $m$ represents four positive constant parameters. We can make some considerations about the values which can be assumed by the parameters present in Eq. (\ref{aeme}):
\begin{itemize}
\item if both $a$ and $m$ are negative defined, then the emergent scenario produces the Big Bang singularity at the infinity paste time, \emph{i.e.} for $t = -\infty$
\item $a_0$ must be a positive defined quantity if we want to have the scale factor of the emergent scenario as a   positive defined quantity
\item $a$  or $m$ must be positive defined  in order to obtain  an expanding model of the Universe
\item $\lambda$ must be a positive defined quantity if we want to avoid singularities (like the Big Rip) at finite time $t$
\end{itemize}
Consequences of this choice are discussed in \cite{debnachatto,ghoshchatto,debna,paul}.\\
The emergent scenario of the Universe in the framework of DE has been taken into account in many recent papers. Ghosh \emph{et al.} \cite{mioghosh} studied the Generalized Second Law of Thermodynamics (GSLT) for the emergent scenario of the Universe for some particular models of $f\left(T\right)$ modified gravity theory. Mukherjee \emph{et al.} \cite{mukhe} studied a general context for an emergent Universe scenario and they derived that the emergent Universe scenarios do not represent isolated solutions but they can occur for different combinations of matter and radiation. del Campo \emph{et al.} \cite{delcampo0} considered the emergent model of scale factor in the framework of a self-interacting Jordan-Brans-Dicke modified gravity theory: they derived that this model is able to lead to a stable past eternal static solution which eventually is able to enter a phase where the stability is broken, which leads to a period of inflation.\\
The first time derivative of the scale factor for the emergent scenario given in Eq. (\ref{aeme}) is:
\begin{eqnarray}
\dot{a}\left(  t \right) = a_0 m \mu  e^{ \mu t}\left[  \lambda + e^{ \mu t  } \right]^{m-1}. \label{dotaeme}
\end{eqnarray}
Using the definition of the scale factor given in Eq. (\ref{aeme}) along with its time derivative given in Eq. (\ref{dotaeme}), we can easily derive that the Hubble parameter $H$ and its first and second derivatives with respect to the cosmic time $t$ are given, respectively, by the following expressions:
\begin{eqnarray}
 H  &=& \frac{\dot{a}}{a}= \frac{e^{\mu t} m \mu }{e^{\mu t}+\lambda }, \label{heme}\\
\dot{H} &=& \frac{\dot{H}}{dt}= \frac{m\lambda \mu^2e^{ \mu t  } }{\left[e^{ \mu t  } + \lambda\right]^2}, \label{dotheme}\\
\ddot{H} &=& \frac{\ddot{H}}{dt^2} =\frac{m\lambda \mu^3e^{ \mu t  }\left( \lambda - e^{ \mu t  }  \right) }{\left[e^{ \mu t  } + \lambda\right]^3}\label{ddoteme}.
\end{eqnarray}

Using the expression of $H$ obtained in Eq. (\ref{heme}) and the expressions of $\Omega_{D_{GO}}$,  $\Omega_{D_{\tiny\circ}}$ and $\Omega_{higher}$, obtained inserting in Eqs.  (\ref{Fractional1GO}), (\ref{Fractional1MH}) and (\ref{Fractional1higher}) the expressions of $\rho_{D_{GO}}$, $\rho_{D_{\tiny\circ}}$ and $\rho_{higher}$ defined in Eqs. (\ref{lgo5-1}), (\ref{mhrde}) and (\ref{lgo5-1higher}) calculated for  $H$, $\dot{H}$ and $\ddot{H}$ given in Eqs.  (\ref{heme}), (\ref{dotheme}) and (\ref{ddoteme}), we derive the following expressions for $\dot{u}_{GO}$, $\dot{u}_{\tiny\circ}$ and $u_{higher}$:
\begin{eqnarray}
 \dot{u}_{GO} &=& 3u_{GO}\left(  1+u_{GO} \right) \left[  \frac{e^{2 t \mu } m^2 \mu }{c^2 \left(e^{t \mu }+\lambda \right) \left(e^{t \mu } m \beta +\alpha  \lambda \right)}  \right]  \times \nonumber \\
  &&\left\{  \left[ \frac{2}{3}u\left( \frac{\alpha - 2\beta}{\beta}  \right)  + \frac{2\alpha}{3\beta}-\frac{4}{3}   \right] \frac{e^{t \mu } m}{c^2 \left(e^{t \mu } m \beta +\alpha  \lambda \right)\left(1+u_{GO}\right)} \right. \nonumber \\
&& \left.  -\frac{2}{3c^2\beta}  -\frac{b^2\left(1+u_{GO}\right)^2}{u_{GO}}\right\},   \label{ugoeme}\\
 \dot{u}_{\tiny\circ} &=&  3u_{\tiny\circ}\left(  1+u_{\tiny\circ} \right) \left\{ \frac{3e^{2\mu t}m^2 \mu \left( \alpha -\beta \right)}{\left( e^{\mu t} + \lambda \right)  \left(  3 e^{t \mu } m \alpha +2 \lambda  \right)}   \right\}\times \nonumber \\
    &&\left\{ \left[\left(\alpha - \frac{4}{3}\right)\frac{3e^{t \mu } m (\alpha -\beta )}{3 e^{t \mu } m \alpha +2 \lambda } \right]+\beta - \alpha  -\frac{b^2\left(1+u_{\tiny\circ}\right)^2}{u_{\tiny\circ}} \right\}, \label{umheme}\\
 \dot{u}_{higher} &=& 3u_{higher}\left(  1+u_{higher} \right) \left[  \frac{e^{2 t \mu } m^2 \mu }{c^2 \left(e^{t \mu }+\lambda \right) \left(e^{t \mu } m \beta +\alpha  \lambda \right)}  \right]  \times \nonumber \\
  &&\left\{  \left[ \frac{2}{3}u\left( \frac{\alpha - 2\beta}{\beta}  \right)  + \frac{2\alpha}{3\beta}-\frac{4}{3}   \right] \frac{e^{t \mu } m}{c^2 \left(e^{t \mu } m \beta +\alpha  \lambda \right)\left(1+u_{higher}\right)} \right. \nonumber \\
&& \left. -\frac{2\varrho \left(e^{t \mu } -\lambda  \right)\lambda}{3c^2\beta\left[ m^2\alpha e^{2t \mu } +e^{t \mu }\left(m\beta - \varrho   \right)\lambda + \varrho \lambda^2 \right]}  -\frac{2}{3c^2\beta}  -\frac{b^2\left(1+u_{higher}\right)^2}{u_{higher}}\right\}.  \label{uhighereme}
\end{eqnarray}

\begin{figure}[ht]
\centering\includegraphics[width=8cm]{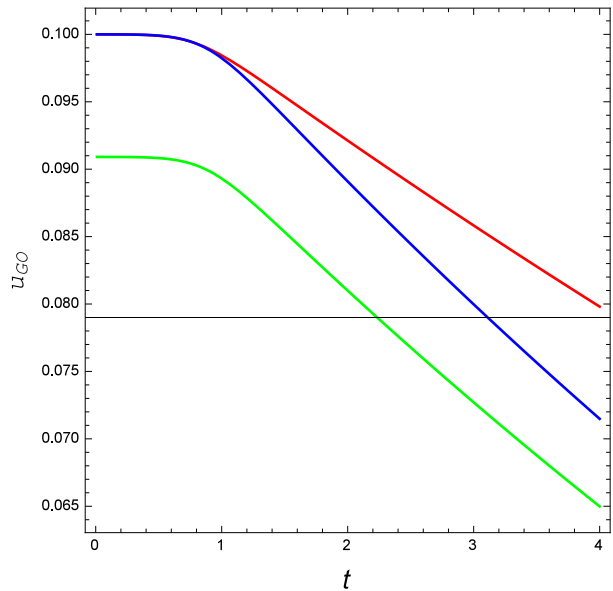}
\caption{Plot of $u_{GO}$ for scale factor emergent scenario. The decreasing pattern indicates that the non-trivial contribution of DE decreases with the evolution of the Universe.}
\label{ugoemergent}
\end{figure}

\begin{figure}[ht]
\centering\includegraphics[width=8cm]{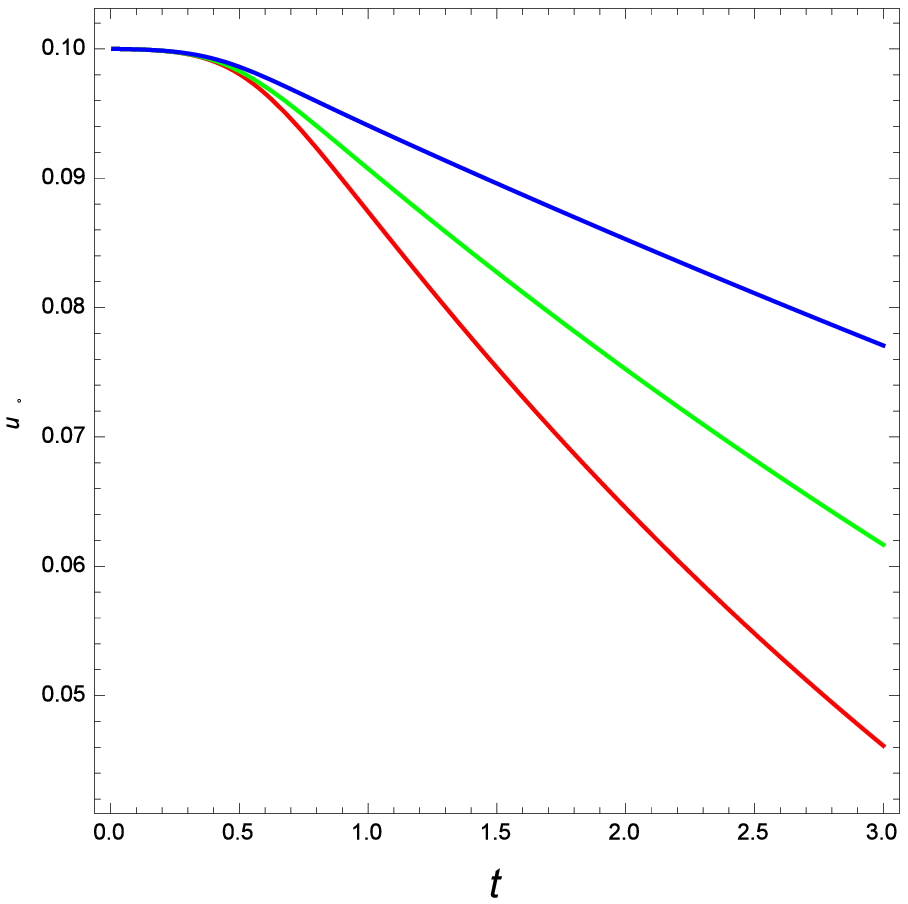}
\caption{Plot of $u_{{\tiny\circ}}$ for scale factor emergent scenario. The decreasing pattern indicates that the non-trivial contribution of DE decreases with the evolution of the Universe.} \label{umhrdeemergent}
\end{figure}

\begin{figure}[ht]
\centering\includegraphics[width=8cm]{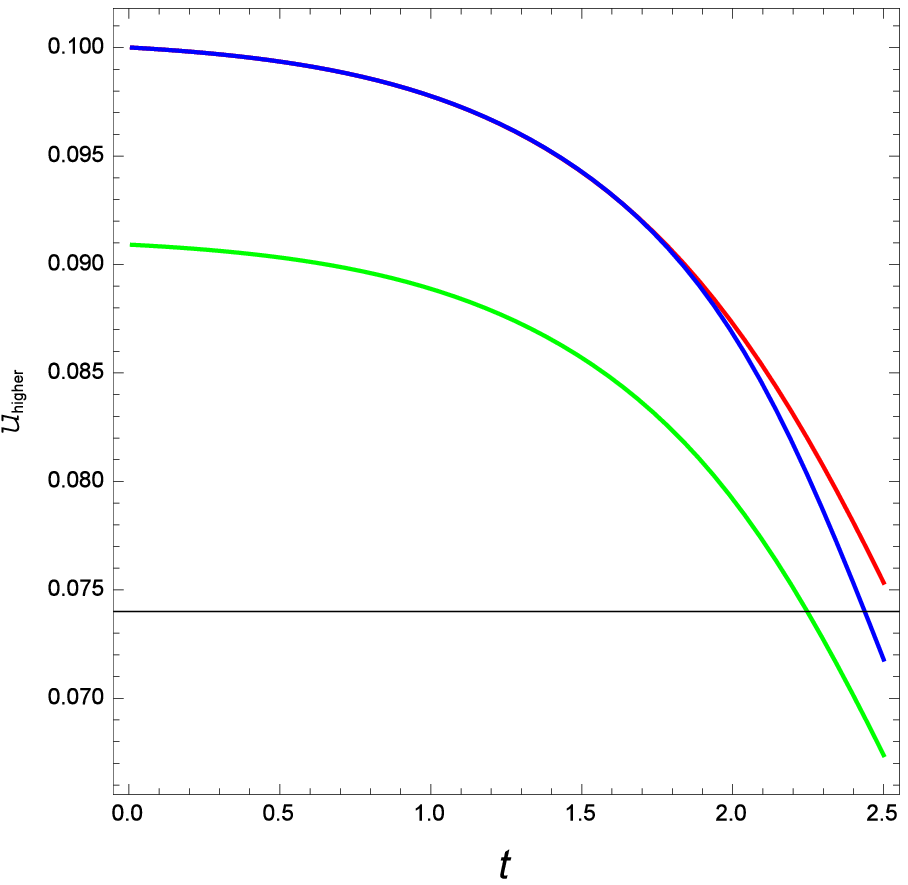}
\caption{Plot of $u_{higher}$ for scale factor emergent scenario. The decreasing pattern indicates that the non-trivial contribution of DE decreases with the evolution of the Universe.}
\label{uhigheremergent}
\end{figure}

As accomplished for the previous model studied, we use a numerical integration in order to obtain the evolutionary forms of $u_{GO}$, $u_{\tiny\circ}$ and $u_{higher}$ and we plot them, respectively, in Figures \ref{ugoemergent}, \ref{umhrdeemergent} \ref{uhigheremergent}.\\
For the to the HDE model with GO cut-off, three different cases have been considered, namely  $\left\{\alpha =4, \beta=8\right\}$ (plotted in red), $\left\{\alpha = 5, \beta=5.8\right\}$     (plotted in green) and $\left\{\alpha =3,\beta=5 \right\}$ (plotted in blue), while the other parameters involved have been chosen as $m=0.03$, $\mu=6$, $\lambda = 5$, $c^2=0.818$ and $b^2=0.025$. We can clearly observe that $u_{GO}$ has a decreasing  behavior for all the three cases considered.\\
For the case corresponding to the MHRDE model, we consider three different cases,   $\left(\alpha=3,\beta=1.5\right)$ (plotted in red),  $\left(\alpha=4,\beta=2.5\right)$ (plotted in green) and $\left(\alpha=6,\beta=4.5\right)$  (plotted in blue) , while the other parameters involved have been chosen as $m=0.03$, $\mu=6$, $\lambda = 5$, $c^2=0.818$ and $b^2=0.025$.  Similarly to $u_{GO}$, $u_{\tiny\circ}$ has a  decreasing behavior for all the three cases considered.\\
For the model proportional to higher time derivatives of the Hubble parameter $H$, we considered three different models, one with $\varrho = 3$ (plotted in red), one with $\varrho = 3.5$ (plotted in green) and one with $\varrho = 4$ (plotted in blue). The other parameters have been chosen as follows: $\alpha = 3.5$, $\beta =3$, $\mu = 1.1$, $\lambda = 5$, $m=5$, $c=0.818$ and $b^2 = 0.025$. We can observe in Figure \ref{uhigheremergent} that $u_{higher}$ has a decreasing behavior for all the cases considered.\\
Therefore, we conclude that we find a decreasing behavior for all the three DE models considered for all the range of values we considered.\\

\begin{figure}[ht]
\centering\includegraphics[width=8cm]{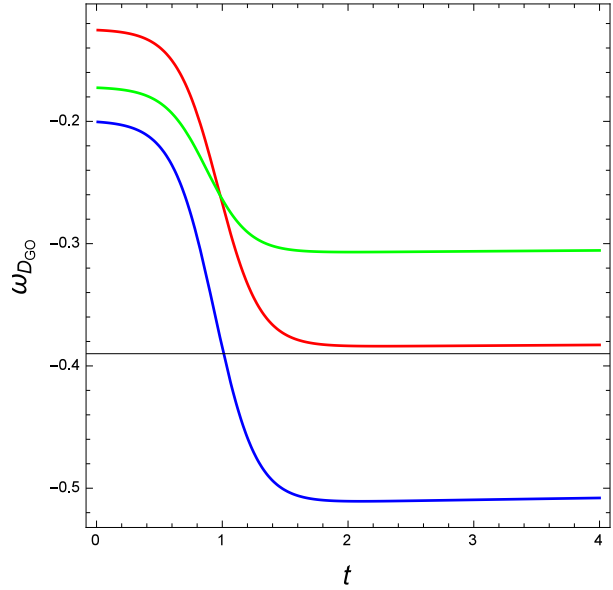}
\caption{Plot of $\omega_{D_{GO}}$ for scale factor emergent scenario. $\omega_{D_{GO}}$ has a decreasing behavior for all the three cases considered.} \label{omegagoemergent}
\end{figure}

\begin{figure}[ht]
\centering\includegraphics[width=8cm]{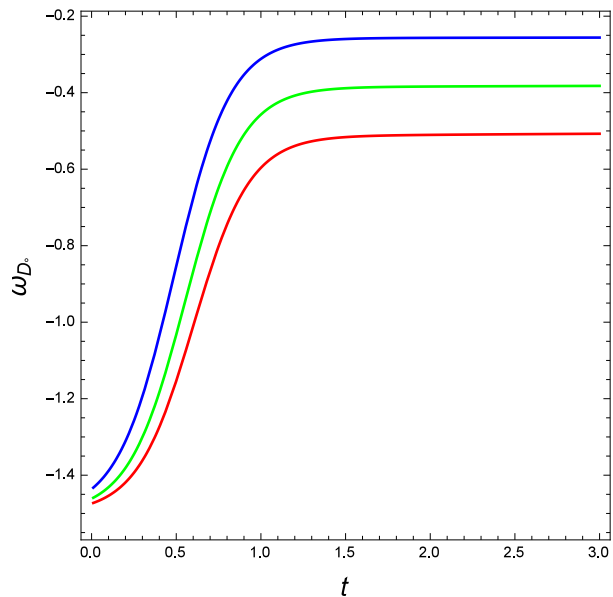}
\caption{Plot of $\omega_{D_{\tiny\circ}}$ for scale factor emergent scenario. $\omega_{D_{\tiny\circ}}$ can go beyond the phantom phase of the Universe in all cases. } \label{omegamhrdemergent}
\end{figure}

\begin{figure}[ht]
\centering\includegraphics[width=8cm]{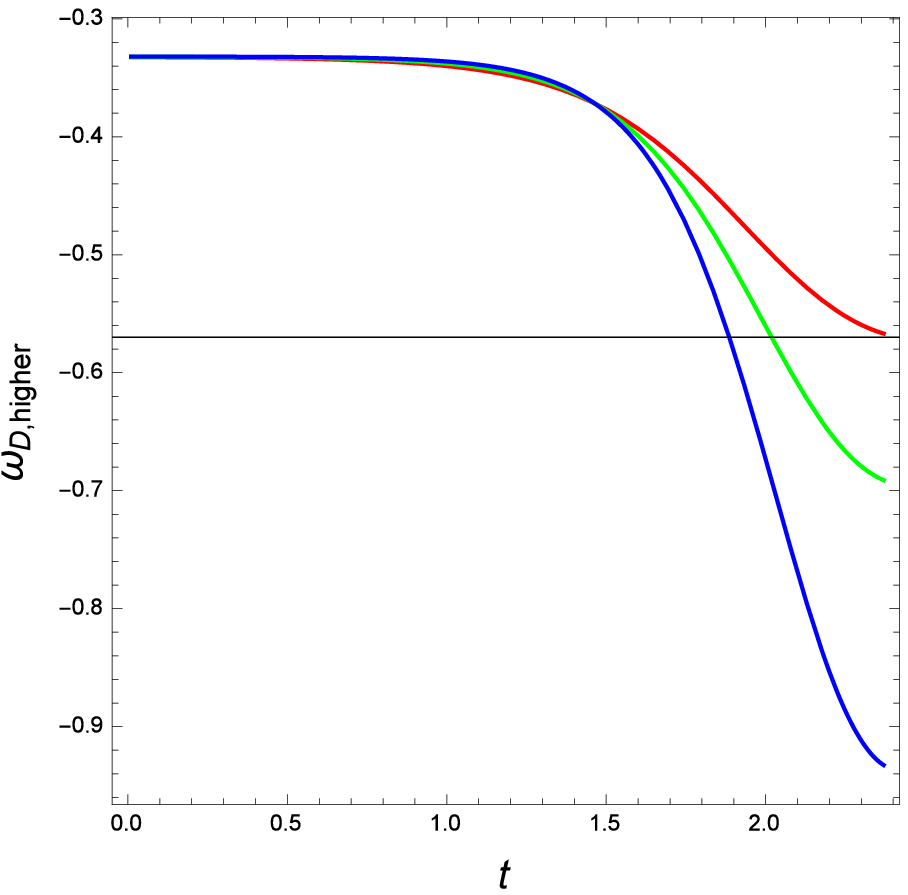}
\caption{Plot of $\omega_{D,higher}$ for scale factor emergent scenario. $\omega_{D,higher}$ has a decreasing behavior for all the three cases considered. Only the case with $\varrho = 4$ (which is plotted in blue) is able to cross $\omega_D =-1$, instead for the other two models we always have  $\omega_D >-1$.} \label{omegahigheremergent}
\end{figure}

Using the expression of $u_{GO}$, $u_{\tiny\circ}$ and $u_{higher}$ obtained, respectively, from Eqs. (\ref{ugoeme}), (\ref{umheme}) and (\ref{uhighereme}) and plotted in Figures \ref{ugoemergent},   \ref{umhrdeemergent} and \ref{uhigheremergent}, we can also plot the  EoS parameters for the three DE models considered in this paper derived in Eqs. (\ref{eosgo1}), (\ref{eosmh1}) and (\ref{eoshigher1}). The EoS parameter of the DE model with GO cut-off $\omega_{D_{GO}}$ has a decreasing behavior, staying always in the region corresponding to $\omega_D>-1$. Moreover, $\omega_{D_{GO}}$ assumes a constant value of $\left[-0.3,-0.5\right]$ (according to the values of the parameters considered) for $t\approx 1.5$.\\
Studying the plot of $\omega_{D_{\tiny\circ}}$, we observe an increasing behavior of the EoS parameter for all the three cases considered. Therefore, we have that $\omega_{D_{\tiny\circ}}$ can go beyond the phantom phase of the Universe in all cases we considered.\\
For the case pertaining to the DE model proportional to $H^2$ and to higher time derivatives of the Hubble parameter $H$, we observe that all the cases considered have a decreasing behavior. Moreover, we observe that only the case with $\varrho = 4$ and plotted in blue can cross the line $\omega_D =-1$, while the other two models always stay in the region $\omega_D >-1$.\\

As done for the power law scale factor studied in the previous Section, we now consider some particular values of the parameters involved. For the DE model with GO cut-off, we study the case corresponding to the Ricci scale, which is recovered for $\alpha = 2$ and $\beta =1$ (plotted in green), and we also consider the case corresponding to $\alpha = 0.8502$ and $\beta = 0.4817$ (plotted in   red). Instead, for the MHRDE model, we consider the case corresponding to the Ricci scale, which is recovered for $\alpha = 4/3$ and $\beta =1$. The values of all the other parameters are taken as the previous cases considered.
\begin{figure}[ht]
\centering\includegraphics[width=8cm]{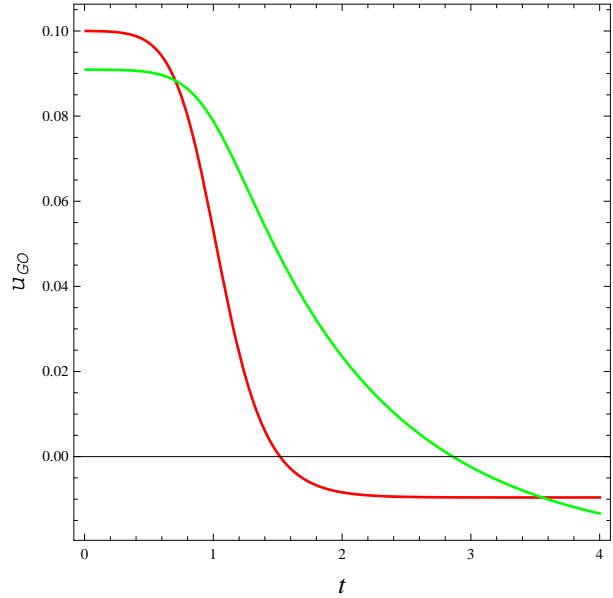}
\caption{Plot of $u_{GO}$ for scale factor emergent scenario against the time $t$ for the limiting cases of $\alpha = 2$ and $\beta =1$ (plotted in   green) and $\alpha = 0.8502$ and $\beta = 0.4817$ (plotted in   red). }
\label{ugoemergentlim}
\end{figure}

\begin{figure}[ht]
\centering\includegraphics[width=8cm]{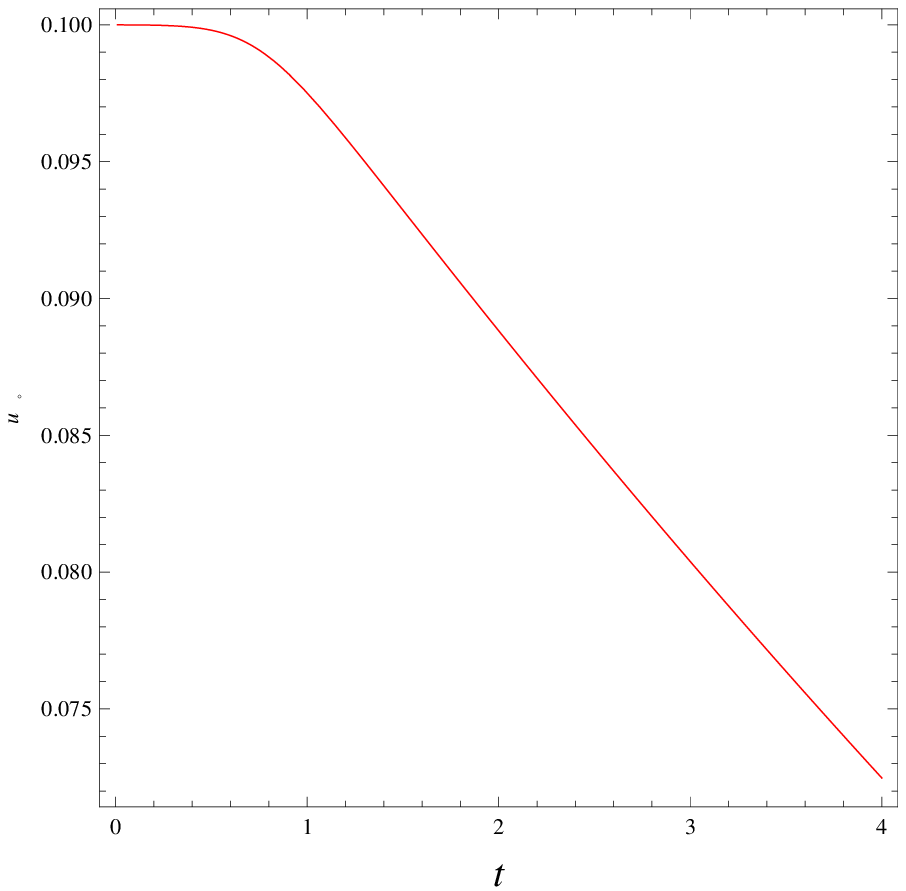}
\caption{Plot of $u_{{\tiny\circ}}$ for scale factor emergent scenario against the time $t$ for the limiting case of  $\alpha = 4/3$ and $\beta =1$. } \label{umhrdeemergentlim}
\end{figure}

\begin{figure}[ht]
\centering\includegraphics[width=8cm]{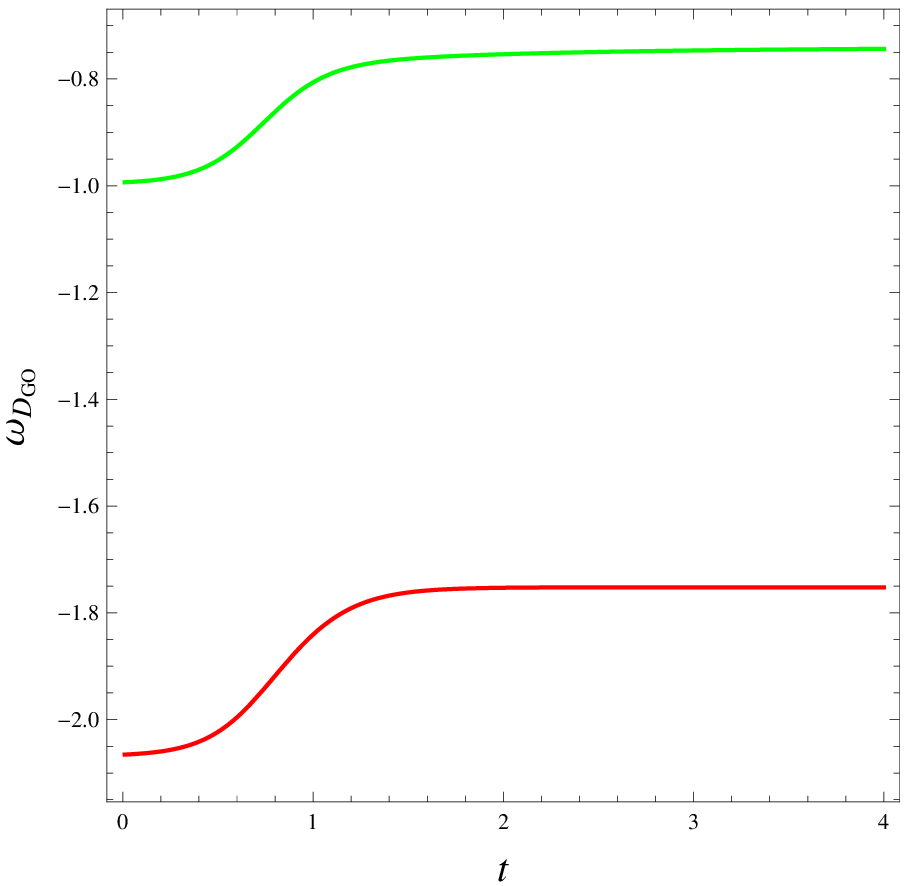}
\caption{Plot of $\omega_{D_{GO}}$ for scale factor emergent scenario against the time $t$ for the limiting cases of $\alpha = 2$ and $\beta =1$ (plotted in   green) and $\alpha = 0.8502$ and $\beta = 0.4817$ (plotted in   red). } \label{omegagoemergentlim}
\end{figure}

\begin{figure}[ht]
\centering\includegraphics[width=8cm]{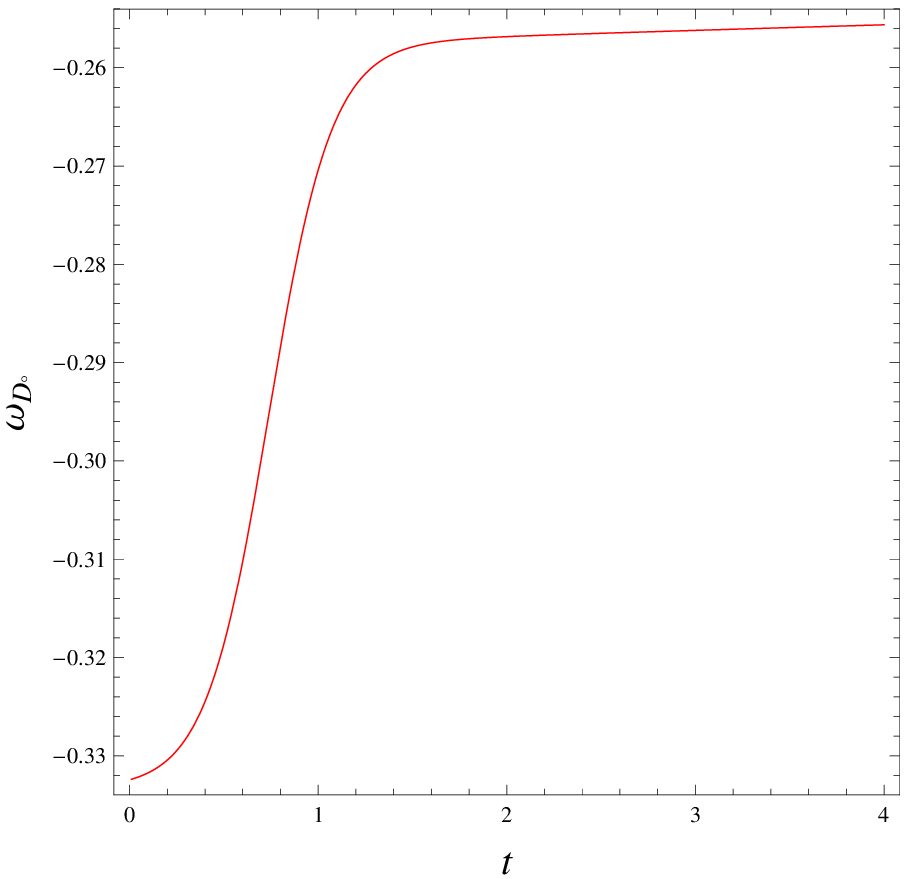}
\caption{Plot of $\omega_{D_{\tiny\circ}}$ for scale factor emergent scenario against the time $t$ for the limiting case of  $\alpha = 4/3$ and $\beta =1$. } \label{omegamhrdemergentlim}
\end{figure}

We can clearly observe in Figure \ref{ugoemergentlim} that $u_{GO}$ has a decreasing behavior for both limiting cases considered. Moreover, for the case with $\alpha = 0.8502$ and $\beta = 0.4817$, $u_{GO}$ starts to assume a constant value for $t \approx 2$. Instead, the EoS parameter $\omega_{D_{GO}}$ has an initial increasing behavior for both case considered, becoming constant for $t \approx 1.4$. Moreover, for the case corresponding to the Ricci scale, we have $\omega_{D_{GO}}$ staying always greater than -1 while for the case with $\alpha = 0.8502$ and $\beta = 0.4817$,   $\omega_{D_{GO}}$ is always lower than -1.\\
For the MHRDE model, we observe that $u_{{\tiny\circ}}$ has a decreasing behavior while the EoS parameter $\omega_{D_{\tiny\circ}}$ start with an increasing behavior and it becomes constant from $t \approx 1.8$, being always greater than -1.

\section{Conclusion}
In this work, we have investigated and studied the effects which are produced by the interaction between a brane Universe and the bulk in which the Universe is embedded. We have assumed that the adiabatic equation for the DM is satisfied, while it is violated for the DE due to the energy exchange between the brane and the bulk. Taking into account the effects of the interaction between a brane Universe and the bulk, we have obtained the EoS parameter for the interacting HDE model with Granda-Oliveros cut-off, the Modified Holographic Ricci DE (MHRDE) model and the DE model proportional to $H^2$ and to higher time derivatives of the Hubble parameter having their energy densities given by $\rho_{D_{GO}}= 3c^2\left( \alpha H^{2}+\beta \dot{H}\right)$,  $\rho_{D_{\tiny\circ}}= \frac{2}{\alpha - \beta} \left(\dot{H} + \frac{3\alpha}{2}H^2    \right)$ and $\rho_{D,higher}= 3c^2\left( \alpha H^{2}+\beta \dot{H} + \varrho \frac{\ddot{H}}{H}\right)$, respectively. Moreover, we must underline that we are considering a flat Universe, then $k=0$.\\
We have considered two choices of scale factor, namely, the power-law  and the emergent ones. The rate of interaction has been taken as $ \Gamma=3b^2 (1+u)H$. We observed that, for the model pertaining to the power law scale factor, the parameter $u$ has an increasing pattern for all the three DE energy density models considered while, for the scale factor pertaining to the emergent case, the parameter $u$ has a decreasing pattern for all the three DE energy density models considered. These observation are valid for all the values of the parameters considered.\\
We have also studied the behavior of the EoS parameter using the reconstructed parameter $u$. We first considered the model with power law scale factor. For the DE model with GO cut-off, we obtained that, for the case corresponding to $\beta = 4.4$,  $\omega_{D_{GO}}$ starts being $>-1$, while with the passing of the time, it asymptotically reaches the point $-1$ and can eventually cross it. For the other two cases considered, we obtained that $\omega_{D_{GO}}$ has a decreasing behavior, being always lower that $-1$. Instead, for the MHRDE model, we obtained that $\omega_{D_{\tiny\circ}}$ has a slowly decreasing behavior for all the three cases considered, staying always greater than $-1$. Moreover,  for the model proportional to higher derivatives of the Hubble parameter $H$, we obtain that $\omega_{D,higher}$ has a decreasing behavior for all the cases considered, staying always in the region $\omega_D>-1$.\\
Considering the case corresponding to the emergent scale factor, we obtained that $\omega_{D_{GO}}$ has a decreasing behavior, staying always in the region $\omega_D>-1$. Furthermore, $\omega_{D_{GO}}$ assumes a constant value in the region $\left[-0.3,-0.5\right]$ for $t\approx 1.5$. Studying the plot of $\omega_{D_{\tiny\circ}}$, we observed an increasing behavior of $\omega_{D_{\tiny\circ}}$  for all the three cases considered. Moreover, we have that $\omega_{D_{\tiny\circ}}$ can go beyond the phantom phase of the Universe in all cases. For the model proportional to higher derivatives of the Hubble parameter $H$, we obtain a decreasing behavior for $\omega_{D,higher}$ for all the cases considered. Furthermore, we have that only the case with $\varrho = 4$ (which is plotted in blue) can cross  the phantom divide line corresponding to $\omega_D =-1$, instead the other two models always stay in the region  $\omega_D >-1$.\\
We also considered the limiting cases corresponding to the Ricci scale for the interacting HDE model with Granda-Oliveros cut-off and for the Modified Holographic Ricci DE (MHRDE) and also the interacting HDE model with Granda-Oliveros cut-off for some particular values of the parameters $\alpha$ and $\beta$ (i.e.   $\alpha = 0.8502$ and $\beta = 0.4817$) derived in a recent work of Wang $\&$ Xu.\\
For the case corresponding to the power law scale factor, we obtained that, for both limiting cases considered, $u_{GO}$ has a decreasing behavior while $\omega_{D_{GO}}$ has a slowly increasing behavior. Moreover, we have that for the case corresponding to the Ricci scale,   $\omega_{D_{GO}}$ stays always greater than the value -1, while for the case with $\alpha = 0.8502$ and $\beta = 0.4817$ it is always lower than -1. For the limiting case of the MHRDE, we observe that $u_{\tiny\circ}$  has an increasing behavior  while $\omega_{D_{\tiny\circ}}$ slowly decreases, being always greater than the value of -1.\\
For the case corresponding to the emergent scale factor, we can obtained that $u_{GO}$ has a decreasing behavior for both limiting cases considered. Moreover, for the case with $\alpha = 0.8502$ and $\beta = 0.4817$, $u_{GO}$ starts to assume a constant value for $t \approx 2$. Instead, we have that the EoS parameter $\omega_{D_{GO}}$ has an initial increasing behavior for both limiting cases considered, becoming constant for $t \approx 1.4$. Moreover, for the case corresponding to the Ricci scale, we have $\omega_{D_{GO}}$ staying always greater than -1 while for the case with $\alpha = 0.8502$ and $\beta = 0.4817$,   $\omega_{D_{GO}}$ is always lower than -1.\\
For the MHRDE model, we observe that $u_{{\tiny\circ}}$ has a decreasing behavior while the EoS parameter $\omega_{D_{\tiny\circ}}$ start with an increasing behavior and it becomes constant from $t \approx 1.8$, being always greater than -1.

\section{Acknowledgement}
SC acknowledges financial support from DST, Govt of India under project grant no. SR/FTP/PS-167/2011.

\end{document}